\documentclass[aps,prb,reprint,showpacs,floatfix]{revtex4-1}  
\usepackage{graphicx}  
\usepackage{bm}        
\usepackage{amssymb}   
\usepackage{amsmath}

\usepackage{color}

\usepackage[extra]{tipa}
\setcitestyle{square,numbers}

\begin{document}

\title{Ground-state and spectral properties of an asymmetric Hubbard ladder}
\author{Anas Abdelwahab}
\author{Eric Jeckelmann}
\affiliation{Institut f\"{u}r Theoretische Physik, Leibniz Universit\"{a}t Hannover, Appelstr.~2, 30167 Hannover, Germany}
\author{Martin Hohenadler}
\affiliation{\mbox{Institut f\"ur Theoretische Physik und Astrophysik,
    Universit\"at W\"urzburg, Am Hubland, 97074 W\"urzburg, Germany}}

\date{\today}

\begin{abstract}
We investigate a ladder system with two inequivalent legs, namely a Hubbard chain and a one-dimensional electron gas. 
Analytical approximations, the density matrix renormalization group method, and continuous-time quantum Monte
Carlo simulations are used to determine ground-state properties, gaps, and spectral functions of this system at half-filling. 
Evidence for the existence of four different phases as a function of the Hubbard interaction and the rung hopping is presented. 
First, a Luttinger liquid exists at very weak interchain hopping. Second, a
Kondo-Mott insulator with spin and charge gaps induced by an effective rung
exchange coupling is found at moderate interchain hopping or strong
Hubbard interaction. Third, a spin-gapped paramagnetic Mott insulator with
incommensurate excitations and pairing of doped charges is observed
at intermediate values of the rung hopping and the interaction.
Fourth, the usual correlated band insulator is recovered for large rung hopping.
We show that the wave numbers of the lowest single-particle excitations are different in each insulating phase.
In particular, the three gapped phases exhibit markedly different spectral functions.
We discuss the relevance of asymmetric two-leg ladder systems 
as models for atomic wires deposited on a substrate.
\end{abstract}


\maketitle

\section{\label{sec:intro}Introduction}

Correlated electrons on ladder lattices have been extensively
investigated in the last two 
decades~\cite{giamarchi,noa94,bal96,sca97,jec98,con05,tsv11,rob12,car13,whi94,dah97,sca02,lav11,shy13}, 
but relatively little attention has
been paid to asymmetric ladders with two inequivalent legs.
The one-dimensional (1D) Kondo-Heisenberg model is the
most studied asymmetric ladder system.
It was used to investigate exotic superconducting correlations
in stripe-ordered high-temperature
superconductors \cite{sik97,zac01b,ber10,dob13}
as well as quantum phase transitions in heavy-fermion materials \cite{eid11}.
Additionally, a two-band Hubbard model on a ladder lattice was the starting
point
of an investigation of pairing mechanisms in strongly repulsive fermion
systems~\cite{alh09}.

In a different context, asymmetric ladder systems have been proposed as models
for linear atomic wires deposited on the surface of a substrate~\cite{spr07,das01}.
In that case, one leg represents the wire while the second leg mimics
those degrees of freedom of the substrate that couple to the wire.
The study of such models provides a first approximation for the influence of the substrate on
hallmarks of 1D physics such as the Peierls
instability~\cite{spr07} and the Luttinger liquid \cite{das01}.
However, this approach has not been pursued systematically until now.

1D electron systems have been studied extensively for more than 60 years~\cite{bae04}.
Well-established theories predict various anomalous properties of strictly 1D electron
systems such as the Peierls instability~\cite{kag82,gruener}, incommensurate 
charge-- and spin-density waves~\cite{gruener}, the dynamical separation of spin
and charge excitations, and the Luttinger liquid behavior of 1D conductors~\cite{giamarchi}.
Experimentally, quasi-1D electron systems have been realized in strongly anisotropic
bulk materials such as Bechgaard salts~\cite{bechgaard} and $\pi$--conjugated polymers~\cite{kiess}.
Experimental and theoretical investigations have both demonstrated that even
a weak coupling between 1D electron systems can play an essential
role for their physical properties~\cite{bae04,giamarchi,bechgaard}.

More recently, quasi-1D electron systems have been realized in 
atomic wires deposited on the surface of a
semiconducting substrate~\cite{spr07,onc08,sni10}.
For instance, it has been claimed that a Peierls metal-insulator transition occurs in indium chains on a silicon 
substrate~\cite{sni10}
and that Luttinger liquid behavior is found in gold chains on a germanium substrate~\cite{blu11}.
However, these claims remain controversial. A fundamental issue is that we have a poor theoretical
knowledge of the  influence of the coupling between wire and substrate. As investigations of 
interacting electrons on three-dimensional lattices with complex 
geometries are extremely difficult, the modeling of wire-substrate
systems by much simpler asymmetric ladders~\cite{spr07,das01} 
appears very promising.

In this paper, we consider a two-leg ladder system made of two inequivalent
legs; one is an interacting electron system described by the 1D Hubbard model
with on-site interaction $U$ and hopping integral $t_{\parallel}$, the other 
is a 1D electron gas (Fermi gas) described by a tight-binding model
with the same $t_{\parallel}$.  The legs are coupled by an interchain (or rung) hopping $t_{\perp}$.
This is the simplest model of a correlated
atomic wire coupled to a noninteracting substrate.
It can also be seen as a special case of the 
general two-band Hubbard model used
to investigate pairing mechanisms \cite{alh09}.
The model is further related to the Kondo-Heisenberg
model~\cite{sik97,zac01b,ber10,dob13,eid11}
because the Hubbard chain at half-filling has the same low-energy spin 
excitations as a Heisenberg chain.
Thus, the asymmetric Hubbard ladder can be seen as a generalization of the
Kondo-Heisenberg model (which corresponds
to a Mott insulator with infinitely large charge gap on the interacting leg) to the case of a Mott insulator 
with a finite gap for charge excitations.

Here, we investigate the model properties for various values of the interaction
$U$ and the rung hopping $t_{\perp}$ in a half-filled ladder, as well as at
low doping away from half-filling.
Ground-state properties, excitation gaps,
and spectral functions are determined accurately
using the density-matrix renormalization group (DMRG) 
technique~\cite{dmrg,dmrg2,dmrg3} and quantum Monte Carlo (QMC) simulations~\cite{rub05}.
Furthermore, approximate analytical methods (perturbation theory and
mean-field approximation) are used to facilitate the interpretation
of the numerical results.
We find that the physics of the half-filled asymmetric ladder is
very rich, with similarities to the Kondo-Heisenberg
model~\cite{sik97,zac01b,ber10,dob13,eid11} and
the half-filled symmetric Hubbard ladder~\cite{giamarchi,noa94,bal96,sca97,jec98}  
(corresponding to a ladder with two identical legs) in certain parameter regimes.
Furthermore, our results confirm that our model is a good starting point 
to investigate an atomic wire deposited on a substrate, but also reveal the
limitations of representing the  substrate by a single chain.

The paper is structured as follows: In Sec.~\ref{sec:model}, we introduce
the model and discuss its properties in limiting cases.
The Hartree-Fock approximation for half-filling is presented in Sec.~\ref{sec:hartree}.
In Sec.~\ref{sec:dmrg}, we discuss our DMRG results for the ground-state
properties and excitation gaps, while the QMC spectral
functions are presented in Sec.~\ref{sec:qmc}. Finally,
Sec.~\ref{sec:conc} contains our conclusions.

\section{\label{sec:model}Model}

\begin{figure}
\includegraphics[width=0.49\textwidth]{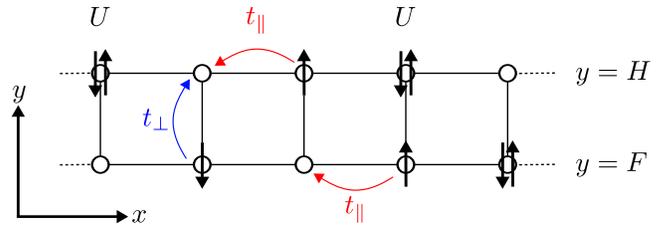}
\caption{\label{fig01} (Color online) The asymmetric Hubbard ladder
  described by Hamiltonian~(\ref{eq:hamiltonian}), with
  intrachain hopping $t_\parallel$ and interchain hopping
  $t_\perp$. On the lower (Fermi, $y=F$) leg, electrons do not interact,
  whereas on the upper (Hubbard, $y=H$) leg, they experience an onsite repulsion $U$.
}
\end{figure}

The Hamiltonian of the asymmetric ladder model takes the form
(see also Fig.~\ref{fig01})
\begin{eqnarray}
H=
&-&t_{\parallel}\sum_{x,y,\sigma} \left ( c_{x+1,y,\sigma}^{\dagger}c_{x,y,\sigma}^{\phantom{\dagger}} 
+ c_{x,y,\sigma}^{\dagger}c_{x+1,y,\sigma}^{\phantom{\dagger}} \right )  \nonumber \\
&-&t_{\perp}\sum_{x,\sigma} \left ( c_{x,\mathrm{F},\sigma}^{\dagger} c_{x,\mathrm{H},\sigma}^{\phantom{\dagger}}  +
c_{x,\mathrm{H},\sigma}^{\dagger}c_{x,\mathrm{F},\sigma}^{\phantom{\dagger}} \right ) \nonumber \\
&+&U \sum_{x} \left ( n_{x,\mathrm{H},\uparrow}-\frac{1}{2} \right ) \left (
n_{x,\mathrm{H},\downarrow}-\frac{1}{2} \right )\,.
\label{eq:hamiltonian}
\end{eqnarray}
Here, $c_{x,y,\sigma}$($c^{\dagger}_{x,y,\sigma}$) is an annihilation (creation)
operator for an electron with spin $\sigma$ on the site with coordinates $(x,y)$
where $y=\mathrm{H}$ (Hubbard leg) or $y=\mathrm{F}$ (Fermi leg) and the rung index $x$ runs 
from $1$ to the ladder length $L$. The corresponding electron number operators
are denoted as $n_{x,y,\sigma} = c_{x,y,\sigma}^{\dagger} c_{x,y,\sigma}^{\phantom{\dagger}}$.
Half-filling corresponds to $N=2L$ electrons on the ladder.
The Hamiltonian is invariant under the particle-hole transformation
$c_{x,y,\sigma} \rightarrow (-1)^x c^{\dagger}_{x,y,\sigma}$.
Therefore, at half-filling its Fermi energy is always equal to 0
and it is sufficient to consider electron fillings $N\geq 2L$.
We will investigate repulsive interactions ($U\geq 0$) only. 
As the canonical gauge transformation $c_{x,\mathrm{H},\sigma} \rightarrow
-c_{x,\mathrm{H},\sigma}$, $c_{x,\mathrm{F},\sigma} \rightarrow c_{x,\mathrm{F},\sigma}$ merely changes the sign of 
$t_{\perp}$, and another canonical gauge transformation 
$c_{x,y,\sigma} \rightarrow (-1)^x c_{x,y,\sigma}$ simply changes the sign of 
$t_{\parallel}$,
we only need to consider $t_{\parallel}\geq 0$ and $t_{\perp}\geq0$.
For our numerical results and figures we use the energy unit
$t_{\parallel}=1$. 

In general, the Hamiltonian~(\ref{eq:hamiltonian}) is not exactly solvable.
However, we can understand some of its properties
by considering limiting cases which are amenable to analytical calculations
or related to known models.

\subsection{Weak interactions\label{sec:weak}}

In the noninteracting case ($U=0$), we recover the well-known 
tight-binding ladder~\cite{giamarchi}.
The Hamiltonian can be diagonalized
using bonding and antibonding rung states. For the single-particle 
eigenstates we obtain a bonding band with dispersion
\begin{equation}
 E_\text{b}(k)=-t_{\perp}-2t_{\parallel}\cos(k)
 \label{eq:disp-b}
\end{equation}
and an antibonding band with dispersion
\begin{equation}
 E_\text{ab}(k)=+t_{\perp}-2t_{\parallel}\cos(k) .
 \label{eq:disp-ab}
\end{equation}
For periodic boundary conditions, the wave numbers $k$ in the first Brillouin zone
$\left [ -\pi,\pi \right ]$ are given
by $k=\frac{2\pi}{L}z$ with an integer $z$ fulfilling $-\frac{L}{2}<z\leq \frac{L}{2}$. 

\begin{figure}[t]
\includegraphics[width=0.29\textwidth]{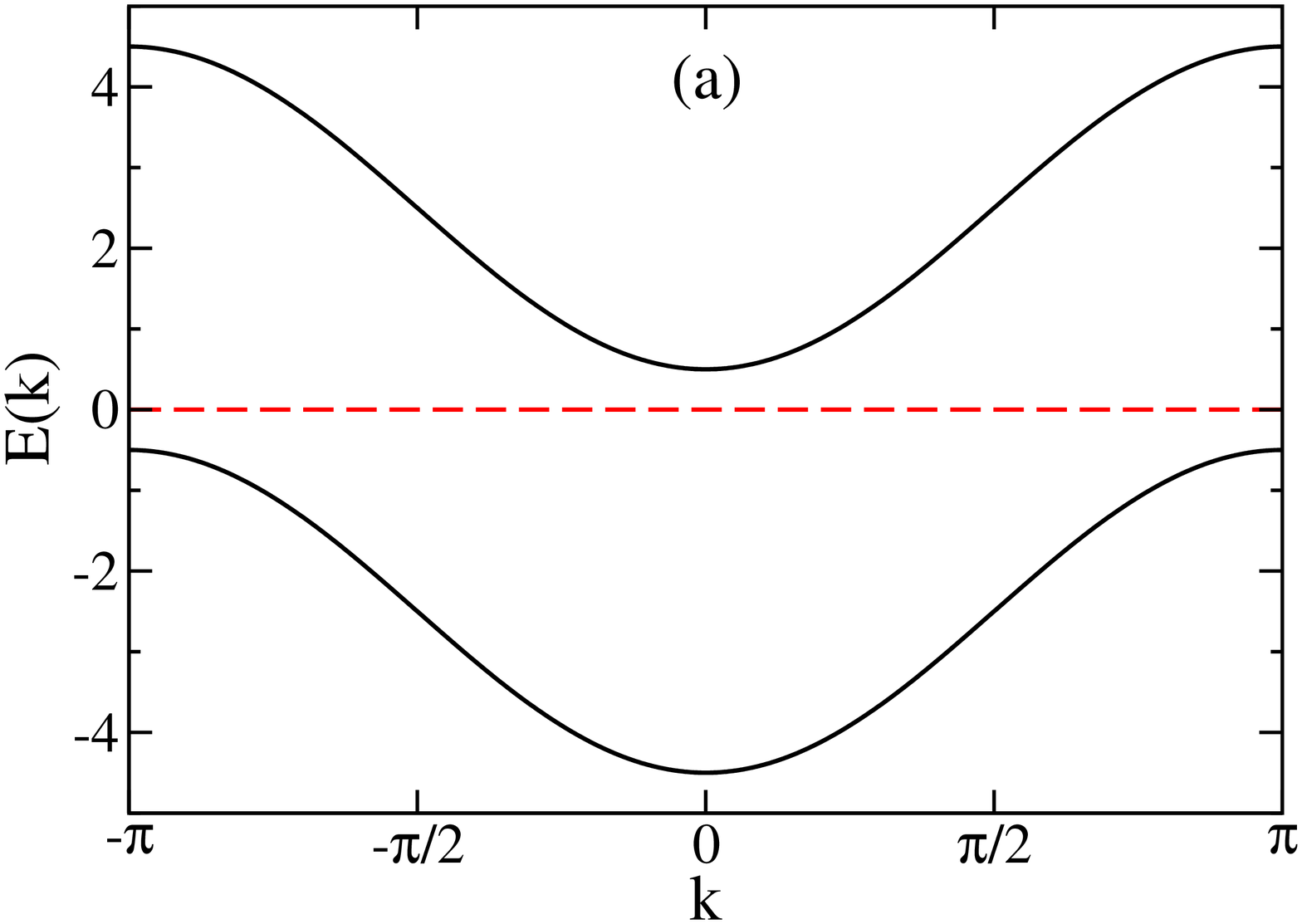}
\includegraphics[width=0.29\textwidth]{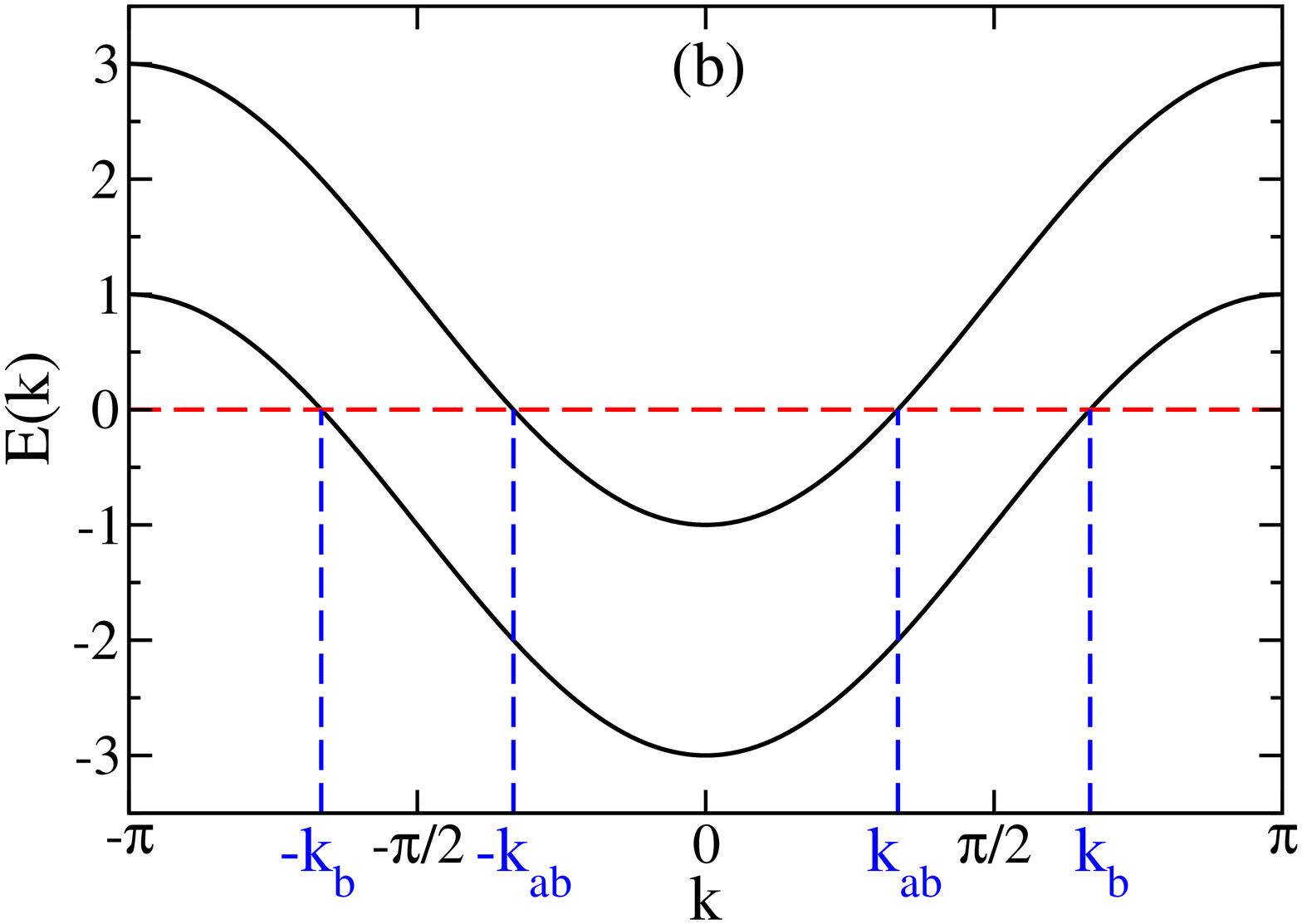}
\includegraphics[width=0.29\textwidth]{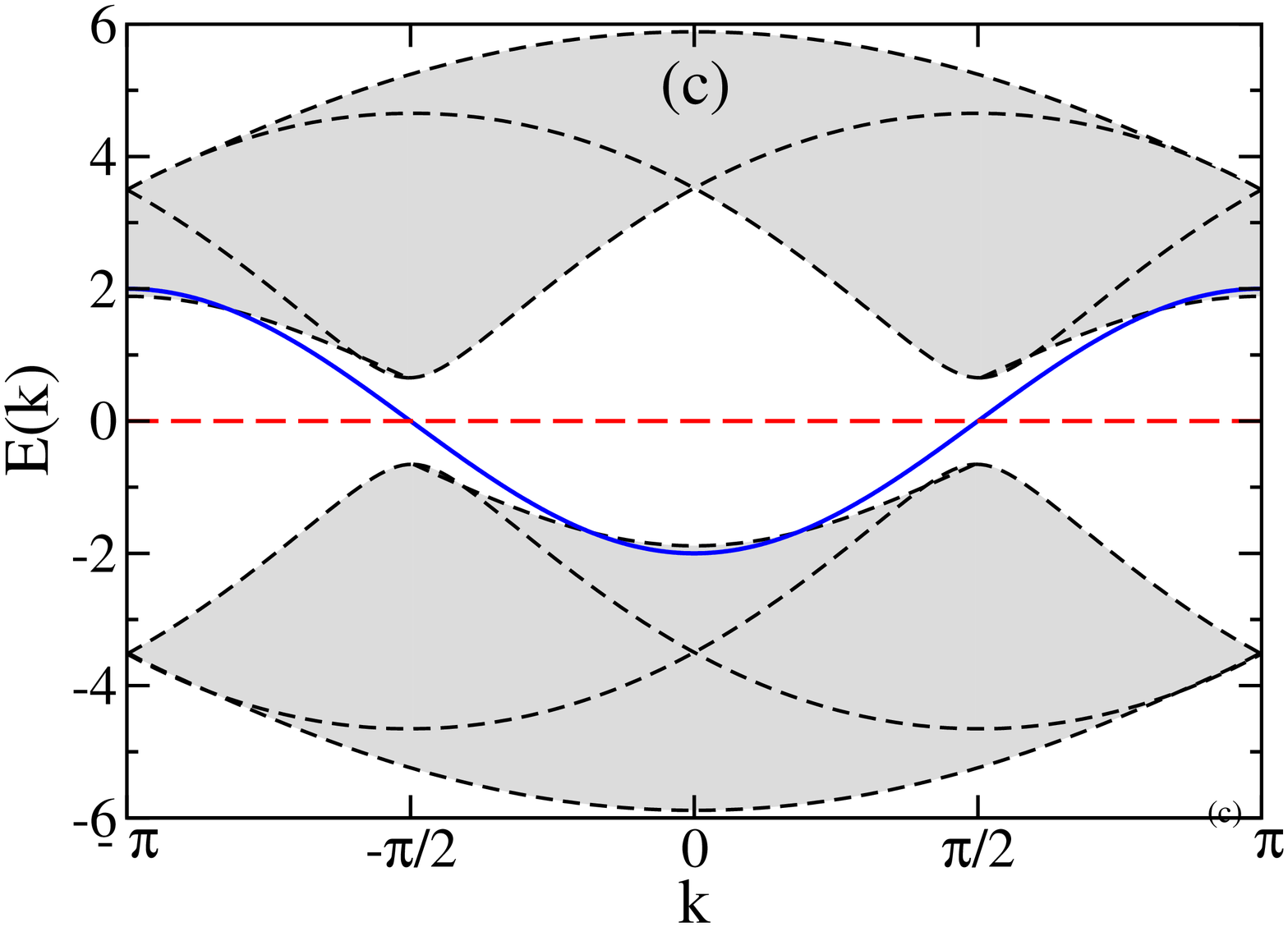}
\caption{\label{fig02} (Color online)
Single-particle dispersions [Eqs.~(\ref{eq:disp-b}) and~(\ref{eq:disp-ab})]
of the noninteracting ladder for (a) $t_{\perp}=2.5 t_{\parallel}$, 
(b) $t_{\perp}=t_{\parallel}$; (b) also shows the four Fermi points $\pm k_{\text{ab}}$ and
$\pm k_{\text{b}}$ defined by Eq.~(\ref{eq:fermi-points}).
(c) Single-particle dispersion of the tight-binding chain (solid blue line) and
single holon-spinon continuum (shaded area) 
of the half-filled Hubbard chain with $U=4t_{\parallel}$ from the Bethe ansatz
solution.
A horizontal dashed line shows the Fermi energy at half-filling in all three figures.
}
\end{figure}

For $t_{\perp}>2t_{\parallel}$ the ladder spectrum has an 
indirect gap 
\begin{equation}
\label{eq:band-gap}
E_{\text{band}} = 2t_{\perp}-4t_{\parallel}
\end{equation}
between the wave numbers $k_\text{b}=\pm \pi$ in the bonding band
and $k_\text{ab}=0$ in the antibonding band, see Fig.~\ref{fig02}(a).
Consequently,
the ladder system is a band insulator at half-filling
while it is metallic with two Fermi points at other band fillings.
Perturbation theory could be used for weak interactions $U \ll
E_{\text{gap}}$, but this case is much easier to analyze in the dimer limit 
(see Sec.~\ref{sec:dimer}).

For $t_{\perp}<2t_{\parallel}$ the ladder spectrum is gapless and
has four perfectly nested Fermi points if the system is at or close to half-filling, see Fig.~\ref{fig02}(b). 
At half-filling the Fermi points $\pm k_{\text{b}}$ 
$\left ( \frac{\pi}{2} < k_{\text{b}} < \pi \right )$  and 
$\pm k_{\text{ab}}$  
$\left ( 0 < k_{\text{ab}} < \frac{\pi}{2} \right )$  
are determined by the equation
\begin{equation}
t_{\perp} = - 2t_{\parallel}\cos(k_{\text{b}}) = 2t_{\parallel}\cos(k_{\text{ab}}) 
\label{eq:fermi-points}
\end{equation}
with the nesting wave number $\pi = k_{\text{b}} + k_{\text{ab}}$.
The case of weak interactions $U \ll t_{\perp}, t_{\parallel}$
could be investigated using sophisticated
field-theoretical approaches (bosonization and the renormalization group),
as done for symmetric ladders~\cite{giamarchi,bal96,con05,tsv11,rob12,car13}.
However, for any finite $U$ the model~(\ref{eq:hamiltonian}) is no longer
symmetric under reflection in the rung direction.
The lower symmetry makes
field-theoretical calculations
much more difficult and, as far as we know, no such calculation has been
carried out successfully for asymmetric Hubbard ladders yet. 
Based on the known results for symmetric
ladders \cite{giamarchi,noa94,bal96,jec98,con05,tsv11,rob12,car13},
we expect that the excitation spectrum of the half-filled
asymmetric ladder becomes fully gapped as soon as $U>0$ 
because the perfect nesting of its Fermi points (with nesting wave number 
$\pi$) allows for umklapp scattering.
The system is then a spin-gapped paramagnetic Mott insulator and its 
lowest single-particle excitations should occur at 
four incommensurate wave numbers $\pm k_{\text{g}}$ and $\pm k'_{\text{g}}$
with $k_{\text{g}}\approx k_{\text{b}}$  and 
$k'_{\text{g}}\approx k_{\text{ab}}$.

\subsection{Strong interactions\label{sec:strong}}

For $t_{\perp} = t_{\parallel}=0$, electrons are localized and 
the ground-state is highly degenerate. At or close to half-filling,
there is exactly one electron on each site
of the Hubbard leg. The other electrons are distributed arbitrarily 
on the Fermi leg.
Using perturbation theory for small but finite hopping terms
($t_{\perp}, t_{\parallel} \ll U$) we find in first order 
that the intrachain hopping 
term $t_{\parallel}$ delocalizes the electrons on the Fermi leg and thus restores
a 1D electron gas with a unique ground state.
The ground state of the Hubbard leg remains unchanged in first order
but
second-order corrections yield the usual antiferromagnetic exchange
coupling $J_{\parallel} = 4t_{\parallel}^2/U$
between electrons localized on nearest-neighbor sites
(and thus an effective 1D Heisenberg model).
The interchain coupling term $t_{\perp}$ yields a constant energy contribution
in second order and thus the legs remain decoupled.
Therefore, it seems that the strong-interaction limit is a special
case of weakly-coupled chains (see Sec.~\ref{sec:chains}).
However, second-order perturbation results are misleading because divergent contributions 
appear at higher orders in $t_{\perp}$. 

The problem at hand is very similar to the single-impurity Anderson
model. Therefore, we can derive an effective Hamiltonian by using a Schrieffer-Wolff
transformation~\cite{sch66}.
Without a hopping term $t_{\parallel}$ in the Hubbard leg,
the asymmetric ladder model~(\ref{eq:hamiltonian}) would be equivalent to a 
1D periodic Anderson model and the
Schrieffer-Wolff transformation (up to the second order) 
would lead to a Kondo lattice
model~\cite{lac79} with an antiferromagnetic exchange interaction $J_{\perp} = 8t_{\perp}^2/U$.
With a hopping term $t_{\parallel} \neq 0$ in both legs, we obtain additional second-order interaction terms:
an antiferromagnetic exchange coupling $J_{\parallel} = 4t_{\parallel}^2/U$ between
nearest-neighbor sites in the Hubbard leg, and next-nearest-neighbor correlated
hopping terms between
Fermi and Hubbard legs of order $t_{\parallel}t_{\perp}/U$. 
Without these correlated hopping terms, the second-order effective Hamiltonian
would be the Kondo-Heisenberg model~\cite{sik97,zac01a,zac01b,ber10,dob13,eid11}.
Hence, the asymmetric ladder with strong Hubbard interaction
can be seen as a generalization of the Kondo-Heisenberg model
to Mott insulators with finite charge gaps.
However, correlated hopping terms are known to be important in the
strong-coupling limit of Hubbard-type models~\cite{esk94}, in particular in
two-leg ladders~\cite{jec98}.
Therefore, contrary to claims in the literature~\cite{ber10}, the
strong-interaction limit of the asymmetric
ladder~(\ref{eq:hamiltonian}) is not exactly equivalent to the
Kondo-Heisenberg model.
However, the Kondo-Heisenberg model could be realized in
the strong-coupling  limit of a simple generalization of 
Hamiltonian~(\ref{eq:hamiltonian}),
for instance by introducing a different intrachain hopping $t^y_{\parallel}$ on each leg.

Nevertheless, for large $U$ we expect the half-filled asymmetric Hubbard ladder 
to exhibit similar low-energy physics as the half-filled Kondo-Heisenberg model
with exchange couplings $J_{\perp} ,J_{\parallel} \ll t_{\parallel}$. 
In the latter model, the rung exchange induces not only a gap for spin excitations
but also for charge excitations in the Fermi leg~\cite{ber10} because of
umklapp scattering associated with perfect nesting of its Fermi points $k_{\text{F}}=\pm \frac{\pi}{2}$.
Additionally, the interaction $U$ is responsible for a large Mott-Hubbard gap on the Hubbard leg
of the asymmetric Hubbard ladder model. We will call this state a Kondo-Mott insulator.

\subsection{Chain limit\label{sec:chains}}

For $t_{\perp}=0$, the model~(\ref{eq:hamiltonian}) reduces to
two independent chains. The first leg corresponds
to a 1D electron gas with a nearest-neighbor tight-binding 
Hamiltonian that can be easily diagonalized.
The second leg is a Hubbard chain which is exactly solvable by the Bethe 
ansatz~\cite{hubbard-book}.
If the ladder system is at or close to half-filling, the Hubbard leg is
exactly half filled because only electronic states of the Fermi leg
are close to the Fermi energy, see Fig.~\ref{fig02}(c). 
Then, the Hubbard leg is a Mott-Hubbard insulator with a
charge gap $E_{\text{H}}$ but gapless spin excitations.
The velocity of spin excitations is smaller than $2t_{\parallel}$ and decreases
with increasing $U/t_{\parallel}$.
The other electrons are on the Fermi leg, which is close to be half filled
and has two Fermi points $k_{\text{F}} \approx \pm \frac{\pi}{2}$ with a Fermi 
velocity $v_{\text{F}} \approx 2t_{\parallel}$.
Therefore, the asymmetric ladder system is metallic, with independent low-energy charge and
spin excitations. Charge excitations 
are localized on the Fermi leg while spin excitations have a lower
velocity on the Hubbard leg than on the Fermi leg.

The interchain hopping term $t_{\perp}$ transfers electrons from one chain to the other and hence
creates excitations with energy larger than $E_{\text{H}}/2$. 
Consequently, for $t_{\perp} \ll E_{\text{H}}$, a perturbative treatment is
possible but merely yields corrections to the eigenenergies because the
ground state is not degenerate. 
However, we expect the interplay of the Hubbard interaction and the 
interchain hopping to induce effective interactions for the electrons in the
Fermi leg, as observed for the strong-interaction limit (see Sec.~\ref{sec:strong}).  The effects of these effective interactions are not known {\it a priori}
but, since a Hubbard chain at half-filling has the same low-energy spin
correlations as a Heisenberg chain, we expect the low-energy physics 
of the weakly coupled chains to be similar to the Kondo-Heisenberg model
with an effective rung exchange coupling $J_{\perp} \propto
t_{\perp}^2/E_{\text{H}} \ll E_{\text{H}}, t_{\parallel}$.  

For weak to moderate interactions $U \alt 4t_{\parallel}$,  the charge gap $E_{\text{H}}$ remains  
small and charge fluctuations between the legs are not negligible.
Thus one cannot assume that the Fermi leg is exactly half filled.
For the Kondo-Heisenberg model away from half-filling, 
various ground states such as Luttinger liquids (with gapless
charge and spin excitations) and spin-gapped phases
with gapless charge excitations have been found \cite{sik97,zac01a,ber10,eid11}.
Nonetheless, we should recover an effective model with a half-filled Fermi
leg for sufficiently large $U$,
as discussed in Sec.~\ref{sec:strong}.
Therefore, various scenarios are possible for the half-filled asymmetric Hubbard ladder in the limit of weakly-coupled chains.
On the one hand, we expect that the ladder system remains
gapless and thus becomes a Luttinger liquid for some range of the parameters $(U,t_{\perp})$.
On the other hand, for large enough $U$, we should recover a Kondo-Mott insulator
with nonzero spin and charge gaps.  Other states are also possible,
as suggested by the studies of the Kondo-Heisenberg model away from
half-filling \cite{sik97,zac01a,ber10,eid11}. In all cases, the lowest
single-particle excitations should remain at the wave numbers given
by the Fermi points of the 1D electron gas, in particular,
$k_{\text{g}} = \pm \frac{\pi}{2}$ for any gapped phase. 
In principle, field theory~\cite{zac01b,ber10,tsv11,rob12}
could be used
to investigate the effects of weak interchain coupling more rigorously.

\subsection{Dimer limit \label{sec:dimer}}

For $t_{\parallel}=0$, we can decompose the Hamiltonian~(\ref{eq:hamiltonian}) 
into a sum of independent two-site Hamiltonians that act on one rung each and
can be easily diagonalized. If the ladder system is half
filled, the ground state corresponds to each rung being occupied by two
electrons that form a spin singlet.
The lowest spin excitation with energy
\begin{equation}
E^{\text{dimer}}_{\text{s}}
 =-\frac{U}{4}+\sqrt{\left ( \frac{U}{4} \right)^{2}+4t_{\perp}^{2}}
\label{eq:dimerEs}
\end{equation}
corresponds to the formation of a triplet on one rung.
The lowest charge excitation with energy
\begin{equation}
\label{dimerEc}
E^{\text{dimer}}_{\text{c}}=-2\sqrt{ \left ( \frac{U}{4} \right )^{2}
+t_{\perp}^{2}}+2\sqrt{\left (\frac{U}{4} \right)^{2}+4t_{\perp}^{2}} 
\end{equation}
corresponds to moving an electron from one rung to the other. 
We note that $E^{\text{dimer}}_{\text{s}} \approx E^{\text{dimer}}_{\text{c}}
\approx 2t_{\perp}$ for $U \ll t_{\perp}$ in agreement with the 
weak-interaction analysis for the band insulating case ($t_{\perp} > 2t_{\parallel}$)
in Sec.~\ref{sec:weak}, while
$E^{\text{dimer}}_{\text{c}} \approx \frac{12t_{\perp}^2}{U}  
> E^{\text{dimer}}_{\text{s}} \approx \frac{8t_{\perp}^2}{U}$ 
for $U \gg t_{\perp}$ in agreement with the rung exchange coupling
deduced for strong interactions in Sec.~\ref{sec:strong}.
If we dope the ladder system away from half-filling by adding electrons,
some of the rungs become occupied by three electrons in the ground state
and both spin and charge gaps drop immediately to zero.

For small but finite $t_{\parallel}$ we can use perturbation theory
as long as $t_{\parallel} \ll E^{\text{dimer}}_{\text{s}},
E^{\text{dimer}}_{\text{c}}$
which corresponds to an energy scale $\sim t_{\perp}$ for weak interactions
($U  \ll t_{\perp}$) and to $\sim t_{\perp}^2/U$ for strong interactions
($U \gg t_{\perp}$). This gives an effective hopping 
$t^{\text{eff}}_{\parallel} \propto t_{\parallel}$ and an effective attractive interaction
$V^{\text{eff}}_{\parallel} \propto t_{\parallel}^2/E^{\text{dimer}}_{\text{c}}$
between nearest-neighbor rungs.
In summary, the half-filled asymmetric ladder in the dimer limit is 
a correlated band insulator for large enough $t_{\perp}/U$.
For large $U/t_{\perp}$, it may be regarded  as a Kondo-Mott insulator
with spin and charge gaps induced by an effective rung exchange coupling,
as discussed in Secs.~\ref{sec:strong} and~\ref{sec:chains}.

\section{\label{sec:hartree}Hartree-Fock approximation}

To gain a better (qualitative) understanding of the
asymmetric ladder model at half-filling, we apply the 
 Hartree-Fock approach for Hubbard-type interactions~\cite{gebhard}
to Hamiltonian~(\ref{eq:hamiltonian}) and obtain the spin-dependent 
single-particle Hamiltonians
\begin{eqnarray}
\label{eq:hf}
H_{\sigma}=&-& t_{\parallel} \sum_{x,y} \left (
c^{\dagger}_{x,y,\sigma}c^{\phantom{\dagger}}_{x+1,y,\sigma}
+  c^{\dagger}_{x+1,y,\sigma}c^{\phantom{\dagger}}_{x,y,\sigma} \right ) 
\nonumber \\
&-& t_{\perp}\sum_{x} \left ( 
c^{\dagger}_{x,\text{H},\sigma}c^{\phantom{\dagger}}_{x,\text{F},\sigma} + 
c^{\dagger}_{x,\text{F},\sigma}c^{\phantom{\dagger}}_{x,\text{H},\sigma}
\right ) 
\nonumber \\
&+& U \sum_{x} n_{x,\text{H},\sigma}\left ( \left \langle n_{x,\text{H},-\sigma} 
\right \rangle - \frac{1}{2}  \right )\,, 
\end{eqnarray}
where the expectation value of the density on the Hubbard leg
$\left \langle n_{x,\text{H},-\sigma} \right \rangle$ must be 
calculated self-consistently
for the ground state of $H_{-\sigma}$.
The Hartree-Fock approximation is a method for weak interactions $U$.

As discussed in Sec.~\ref{sec:weak}, the Fermi points are perfectly nested
by an interband wave number $k=\pi$ at half-filling and for $t_{\perp} <
2t_{\parallel}$.
Therefore, the most probable symmetry breaking is an antiferromagnetic
spin-density wave
\begin{equation}
 \langle n_{x,\text{H},\sigma}\rangle= \frac{1}{2}+ \sigma 
 (-1)^{x}\frac{m_\text{H}}{2}
\end{equation}
with the staggered magnetization (per site) of the Hubbard leg, $m_\text{H}$, as the order parameter.
Consequently, the unit cell of the effective Hamiltonian~(\ref{eq:hf}) is twice as large 
as that of the original Hamiltonian~(\ref{eq:hamiltonian}) in the leg direction
and contains four sites. 
According to Bloch's theorem, the single-particle Hamiltonians~(\ref{eq:hf})
can be diagonalized by a canonical transformation of the form
\begin{eqnarray}
 \label{trans}
 d_{k,n,\sigma}^{\dagger} &=& \frac{1}{\sqrt{L}}\sum_{x} e^{ikx} 
 \left \{ 
 \left [ u_{kn\sigma}+(-1)^{x}v_{kn\sigma} \right ] c_{x,H,\sigma}^{\dagger} 
 \right . \nonumber \\
 &&\hspace*{5em}+\left . 
 \left [ s_{kn\sigma}+(-1)^{x}t_{kn\sigma} \right ] c_{x,F,\sigma}^{\dagger} 
 \right \}
\end{eqnarray}
with the normalization condition
\begin{equation}
|u_{kn\sigma}|^{2}+|v_{kn\sigma}|^{2}+|s_{kn\sigma}|^{2}+|t_{kn\sigma}|^{2}=1
\end{equation}
and a wave number $k$ in a reduced Brillouin zone 
$\left [ -\frac{\pi}{2},\frac{\pi}{2} \right ]$, i.e., $k=\frac{2\pi}{L}z$
with $-\frac{L}{4}<z\leq\frac{L}{4}$. The index $n=1,2,3,4$ numbers
the four bands.
The four components
$\bm{V}_{kn\sigma} = 
(u_{kn\sigma}, v_{kn\sigma}, s_{kn\sigma},t_{kn\sigma})$ are the solutions of the four-dimensional eigenvalue problem
\begin{equation}
\label{eq:HF1}
H_{k\sigma} \bm{V}_{kn\sigma} = \epsilon_{n\sigma}(k) \bm{V}_{kn\sigma}
\end{equation}
with the Hamiltonian matrix $H_{k\sigma}$ given by
\begin{eqnarray*}
\label{eq:HF2}
 & \begin{pmatrix}&
  \begin{array}{cccc}
  -2t_\parallel\cos(k) & -\frac{1}{2}\sigma Um_{H} & -t_{\perp} & 0 \\
  -\frac{1}{2}\sigma Um_{H} & +2t_\parallel\cos(k) & 0 & -t_{\perp} \\
  -t_{\perp} & 0 & -2t_\parallel\cos(k) & 0 \\
  0 & -t_{\perp} & 0 & +2t_\parallel\cos(k)
  \end{array}
  \end{pmatrix}
\end{eqnarray*}
and the single-particle (Hartree-Fock) eigenenergy $\epsilon_{n\sigma}(k)$.
The staggered magnetization is given by 
\begin{equation}
\label{eq:HF3}
m_{\text{H}} = \sigma \frac{4}{L} \sum_{n=1}^2 \sum_k 
u_{kn\sigma}  v_{kn\sigma} \,,
\end{equation}
where the first sum runs over the lowest two bands only.
Equations~(\ref{eq:HF1}) and~(\ref{eq:HF3}) constitute
a self-consistency problem which can be easily solved numerically.

\begin{figure}
\includegraphics[width=0.39\textwidth]{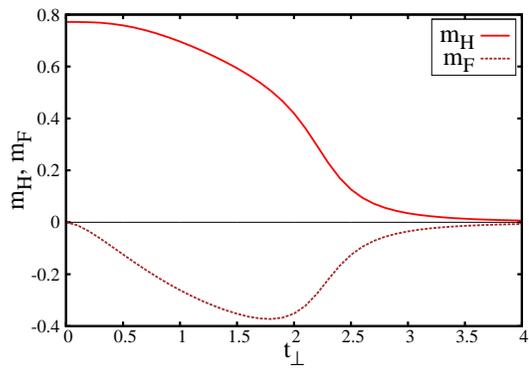}
\caption{\label{fig03} (Color online)
Staggered magnetization of the Hubbard leg ($m_{\text{H}}$) and 
Fermi leg ($m_{\text{F}}$) for $U=4t_{\parallel}$ 
in the Hartree-Fock approximation 
as a function of the rung hopping $t_{\perp}$.
}
\end{figure}

As expected for a 1D system with a perfect nesting of the Fermi
points, we find a broken-symmetry solution $m_{\text{H}}\neq 0$ for any $U > 0$ 
if $t_{\perp} < 2t_{\parallel}$. 
Furthermore, this staggered magnetization seems to remain stable even for
larger $t_{\perp}$
(at least up to $4t_{\parallel}$) although $m_{\text{H}}$ becomes quite small.
However, the long-range antiferromagnetic order 
is an artifact of the mean-field approximation since the continuous $SU(2)$ spin
symmetry can not be spontaneously broken in one
dimension~\cite{giamarchi,gebhard}.
In Fig.~\ref{fig03} we show the self-consistent order parameter $m_{\text{H}}$
obtained for $U=4t_{\parallel}$ as a function of the rung hopping
$t_{\perp}$. (Qualitatively similar results are found for other values of $U$.)
As expected, $m_{\text{H}}$ approaches the value obtained
for the 1D Hubbard model~\cite{gebhard} 
for $t_{\perp} \rightarrow 0$ and
its absolute value decreases monotonically with increasing $t_{\perp}$. 
Although there is no direct electron-electron interaction on the
Fermi leg, the coupling to the Hubbard leg induces an antiferromagnetic
spin-density wave.
The corresponding staggered magnetization, 
\begin{equation}
m_{\text{F}} = (-1)^x \langle n_{x,\text{F},\uparrow} - n_{x,\text{F},\downarrow} \rangle\,,
\end{equation}
is also shown in Fig.~\ref{fig03}. We see that $m_{\text{F}}$ is not a monotonic
function of the interchain coupling $t_{\perp}$.
It vanishes for $t_{\perp}=0$ because the Fermi leg is just an 
independent electron gas in that case (see Sec.~\ref{sec:chains}).
The initial increase of $|m_{\text{F}}|$ with $t_{\perp}$ reflects
the enhanced hybridization of electronic states on the two legs
while the final decrease mirrors the diminution of the antiferromagnetic
correlations in the Hubbard leg.
Note that $m_{\text{H}}$ and $m_{\text{F}}$ have opposite signs
because of the antiferromagnetic correlations between electrons on the 
same rung.

The dispersion of the Hartree-Fock eigenenergies can be calculated 
analytically for a given $m_{\text{H}}$. It has the form
$\epsilon_{n\sigma}(k) = \pm \sqrt{a(k)\pm \sqrt{b(k)}}$ with 
\begin{eqnarray*}
a(k) & = & 
\frac{1}{2} \left ( \frac{Um_{\text{H}}}{2} \right )^2 
+ \left [2t_{\parallel} \cos(k) \right ]^2 + t_{\perp}^2\,,    \\
b(k) & = &
\frac{1}{4} \left ( \frac{Um_{\text{H}}}{2} \right )^4 
+4t_{\perp}^2 \left [2t_{\parallel} \cos(k) \right ]^2
+ \left ( \frac{Um_{\text{H}}}{2} \right )^2 t_{\perp}^2 .
\end{eqnarray*}
The four possible combinations of signs correspond to the four bands
$\epsilon_{n\sigma}(k)$, $n=1,2,3,4$.
(Note that the bands are identical for $\sigma=\pm 1$.)

\begin{figure}
\includegraphics[width=0.39\textwidth]{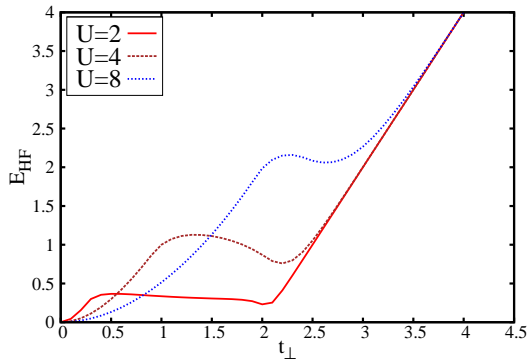}
\caption{\label{fig04} (Color online)
Hartree-Fock gap $E_{\text{HF}}$ for
$U/t_{\parallel}=2,4$ and $8$
as a function of the rung hopping $t_{\perp}$.
}
\end{figure}

\begin{figure}
\includegraphics[width=0.39\textwidth]{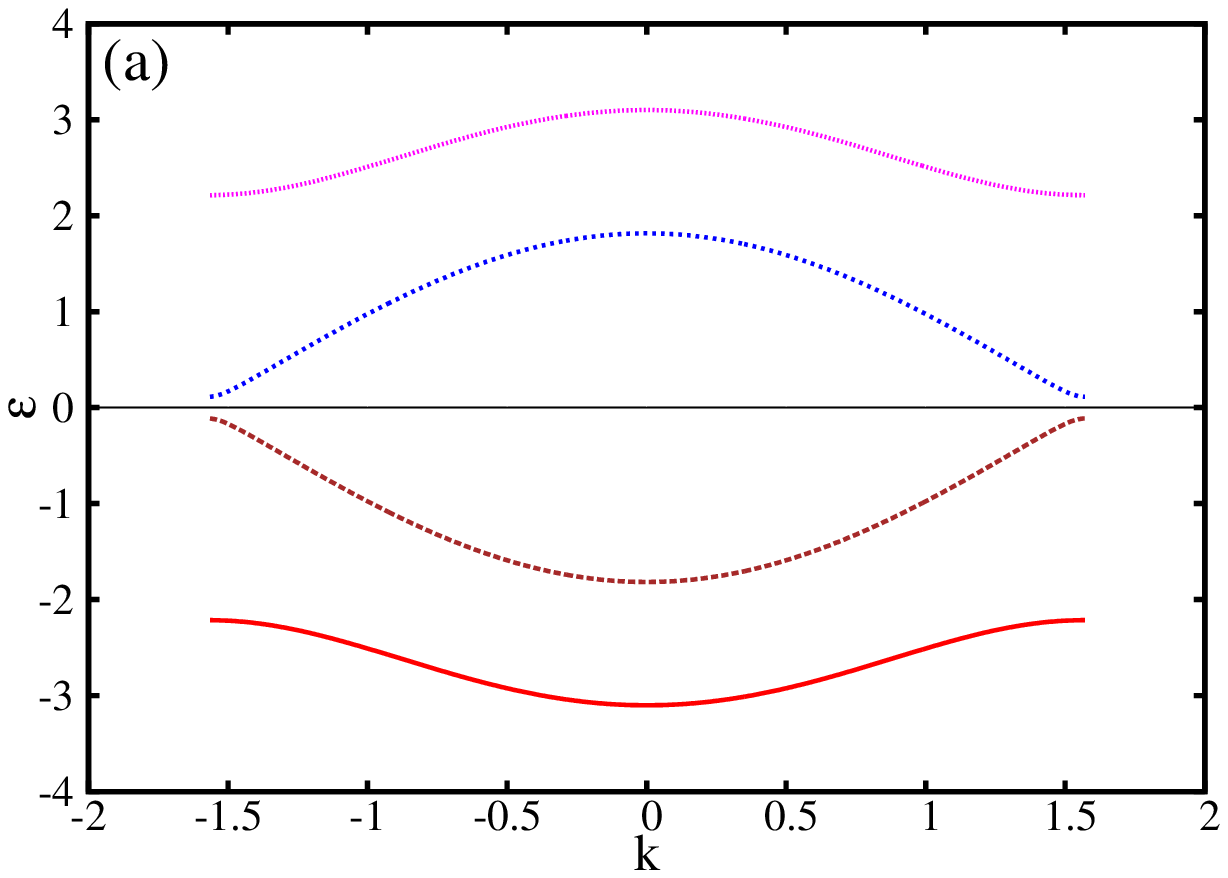}
\includegraphics[width=0.39\textwidth]{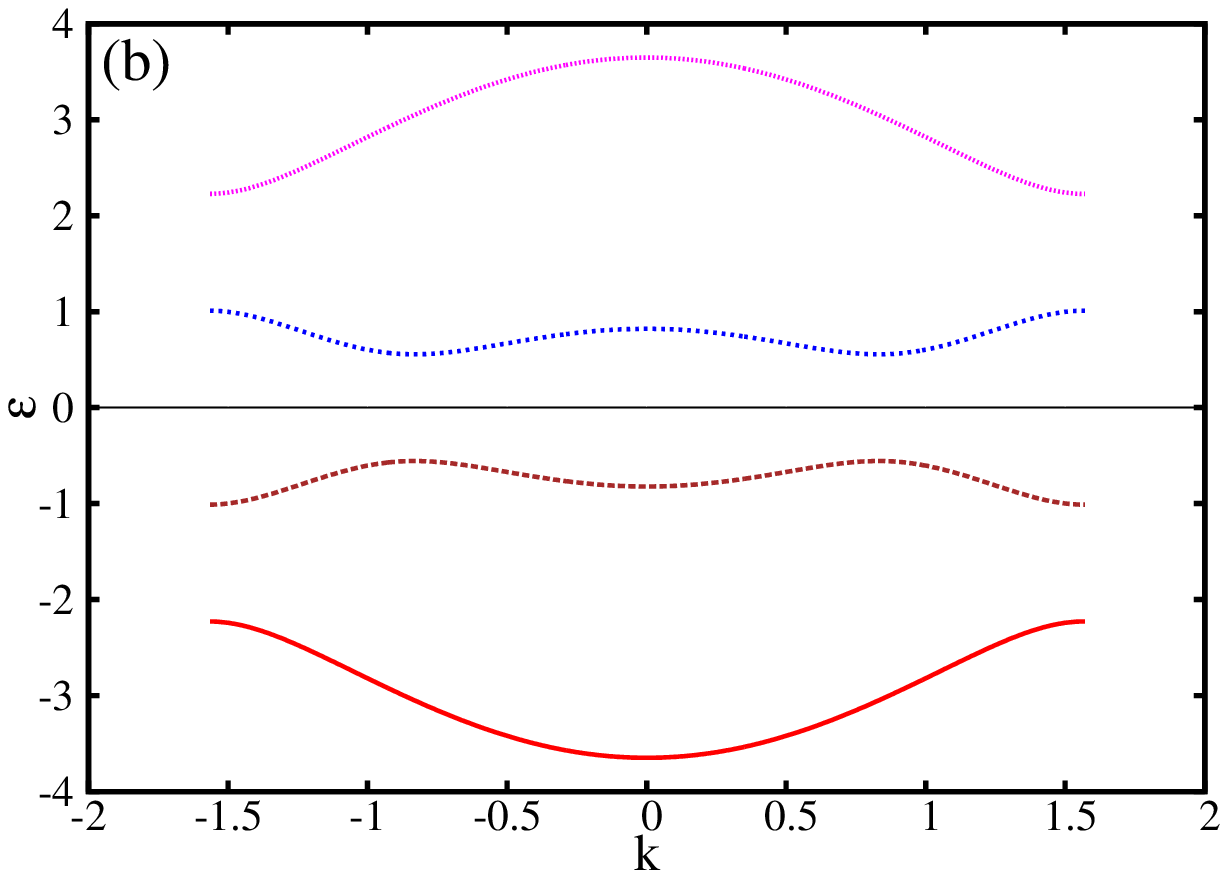}
\includegraphics[width=0.39\textwidth]{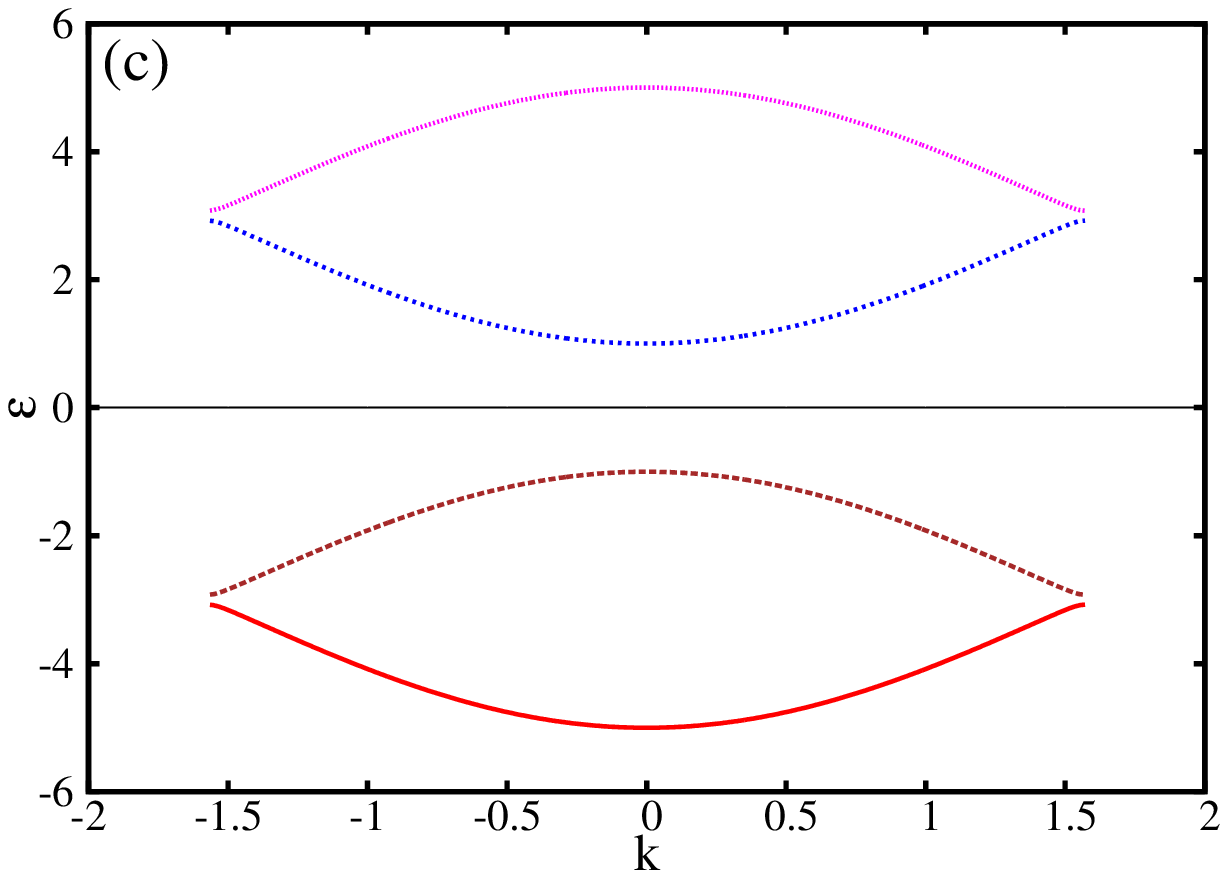}
\caption{\label{fig05} (Color online)
The four Hartree-Fock bands $\epsilon_{n\sigma}(k)$ for $U=5t_{\parallel}$ 
with (a) weak ($t_{\perp}=0.5t_{\parallel}$),
(b) intermediate ($t_{\perp}=1.5t_{\parallel}$), and 
(c) strong ($t_{\perp}=3t_{\parallel}$) rung hopping.}
\end{figure}

The Hartree-Fock gap $E_{\text{HF}}$ is defined as the lowest excitation energy
when the Hartree-Fock bands are half filled, i.e., as
the energy difference between the
lowest state in the third-lowest band and the highest state in the second-lowest band. As expected, this gap vanishes if $U=0$ or $t_{\perp}=0$.
If both couplings are finite, however, we find that the Hartree-Fock
gap is always larger than zero.
The gap has a surprisingly complex dependence on the
interaction strength and the rung hopping, as 
illustrated in Fig.~\ref{fig04}.
We observe three different regions as a function of $t_{\perp}$.
First, the gap is small but increases
rapidly with $t_{\perp}$, then it reaches a local maximum at intermediate
values of $t_{\perp}$ and decreases slowly until it reaches a local minimum
at some value  $t_{\perp} > 2t_{\parallel}$. Finally, it increases
linearly with $t_{\perp}$ at large values of $t_{\perp}$.
The behavior at large $t_\perp$ is easy to understand 
from the discussion of the noninteracting (Sec.~\ref{sec:weak}) and 
dimer limits (Sec.~\ref{sec:dimer}). Indeed, we see that for large $t_{\perp}$
the Hartree-Fock gap approaches the band gap given by Eq.~(\ref{eq:band-gap}).
In this region, the Hartree-Fock solution 
can be regarded as a band insulator with a weak, incidental 
antiferromagnetic ordering. In the other two regions, however, 
the antiferromagnetic ordering is responsible for the gap opening.
These Hartree-Fock solutions describe antiferromagnetic Mott 
insulators~\cite{gebhard}.
For a weak rung hopping the Hartree-Fock gap increases 
systematically with $U$. This case is related to the 
spin-density-wave insulator with  modulation $2k_F=\pi$
which is found in the Hartree-Fock approximation
for 1D half-filled Hubbard-type models.
Note that the extent of the intermediate region in terms of $t_\perp$
decreases upon increasing the interaction $U$.

\begin{figure}
\includegraphics[width=0.39\textwidth]{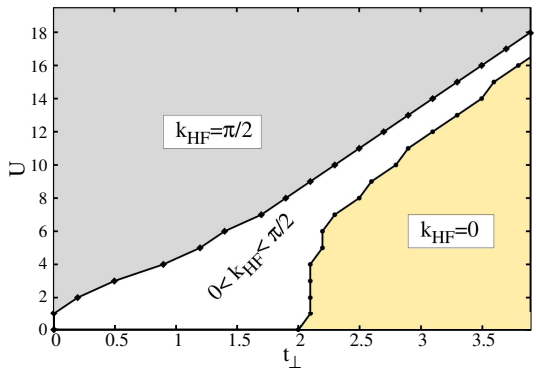}
\caption{\label{fig06} (Color online)
Hartree-Fock ``phase diagram'' in the $\left ( U,t_{\perp} \right )$ plane
with three different regions.
The lowest single-particle excitations have wave numbers $k_{\text{HF}}$ 
at the edges of the reduced Brillouin zone 
$\left ( k_{\text{HF}}=\pm \frac{\pi}{2} \right )$,
at its center $\left ( k_{\text{HF}}=0 \right )$, and at incommensurate values 
$0 < \left | k_{\text{HF}} \right | < \frac{\pi}{2}$, respectively.
}
\end{figure}

The qualitative difference between the first two regions
(weak to moderate rung hopping)
is revealed by studying the features of the Hartree-Fock dispersions $\epsilon_{n\sigma}(k)$.
They are shown in Fig.~\ref{fig05} for a self-consistent
staggered magnetization $m_{\text{H}}$ at $U=5t_{\parallel}$.
For a weak rung hopping [see Fig.~\ref{fig05}(a)]
the lowest single-particle excitations are located at the edge
of the reduced Brillouin zone 
$\left ( k_{\text{HF}} =\pm\frac{\pi}{2} \right )$,
in agreement with the analysis of weakly-coupled chains 
in Sec.~\ref{sec:chains}. Figure~\ref{fig05}(b) shows
that the lowest excitations correspond to single-particle
states with incommensurate wave numbers
$k_{\text{HF}}$
in the intermediate regime in agreement with the analysis
of the case $t_{\perp} < 2t_{\parallel}$ and weak interaction $U$ in
Sec.~\ref{sec:weak}. 
The wave number $k_{\text{HF}}$ determined
from the Hartree-Fock solution shifts progressively
from the edges of the reduced Brillouin zone 
$\left ( k_{\text{HF}}= \pm \frac{\pi}{2} \right )$
to its center ($k_{\text{HF}}=0$) with increasing $t_{\perp}$,
in qualitative agreement with the incommensurate wave number 
given by Eq.~(\ref{eq:fermi-points}).
Finally, for a strong rung hopping $t_{\perp}$
[see Fig.~\ref{fig05}(c)],
the lowest excitations are localized in the center of the reduced Brillouin
zone. This result also agrees with the analysis of the case 
$t_{\perp} > 2t_{\parallel}$ and weak interaction $U$ in Sec.~\ref{sec:weak}.
The indirect gap between $k_{\text{g}}=\pm\pi$ and $k'_g=0$ found there 
[see Fig.~\ref{fig02}(a)]
becomes a direct gap at $k_{\text{HF}}=0$ in the Hartree-Fock approximation
because of the folding of the Brillouin zone.
Finally, the HF ``phase diagram'' in Fig.~\ref{fig06} shows that
all three cases are found over a finite range of the parameters
($U,t_{\perp}$).

\section{\label{sec:dmrg} Ground-state properties and excitation gaps}

\subsection{DMRG method}

To obtain reliable results for the asymmetric ladder 
Hamiltonian~(\ref{eq:hamiltonian}) at finite $U$ and $t_{\perp}$, 
we use the DMRG method \cite{dmrg,dmrg2,dmrg3}, which has previously been 
applied to symmetric~\cite{dmrg2,noa94,whi94,jec98} and asymmetric 
two-leg ladders~\cite{sik97,alh09,ber10,eid11}.
Here, the ground-state properties of Hamiltonian~(\ref{eq:hamiltonian}) 
are calculated using the finite-system DMRG algorithm on lattices with up 
to $L=200$ rungs (400 sites) and open boundary conditions. 
Up to $m=3072$ density-matrix eigenstates were kept, yielding discarded weights
smaller than $10^{-6}$.
Truncation errors were investigated systematically by keeping variable numbers
of density-matrix eigenstates and  
ground-state energies were extrapolated to the limit of vanishing discarded
weights. The resulting error estimates for gaps are shown in the figures
when they are larger than the symbol sizes. 
We were able to reach a sufficient accuracy for the lowest eigenenergies 
for all parameters but for weakly-interacting, weakly-coupled chains
with $U \leq 4t_{\parallel}$ and $t_{\perp} < 2t_{\parallel}$.
As usual with variational approaches, the accuracy is lower for 
other observables
(density profiles, correlation functions).
In some cases, irregular density profiles and correlation functions
demonstrate that the DMRG calculation has not fully converged
because of (quasi-) degenerate low-lying eigenstates. 
The relevant cases are discussed below together with our results.

\subsection{Excitation energies}

\begin{figure}
\includegraphics[width=0.39\textwidth]{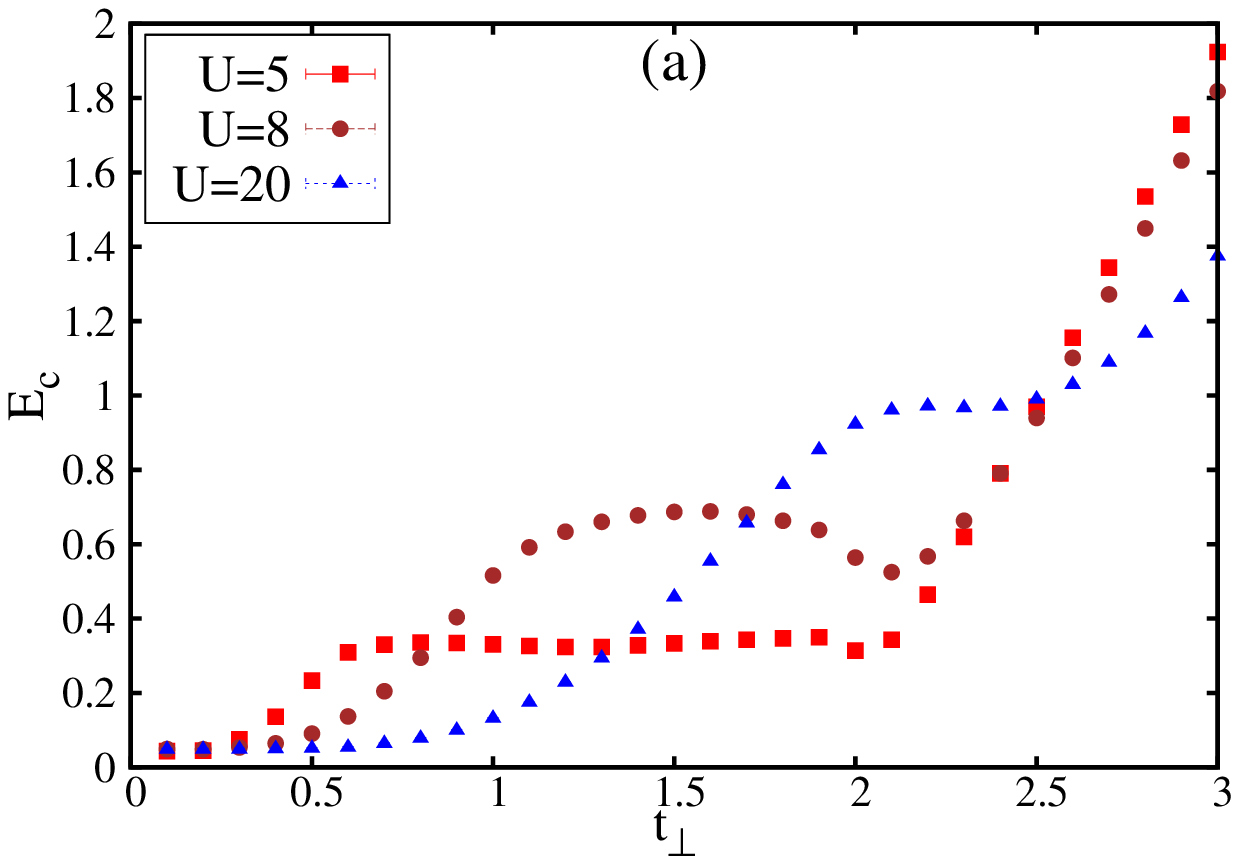}
\includegraphics[width=0.39\textwidth]{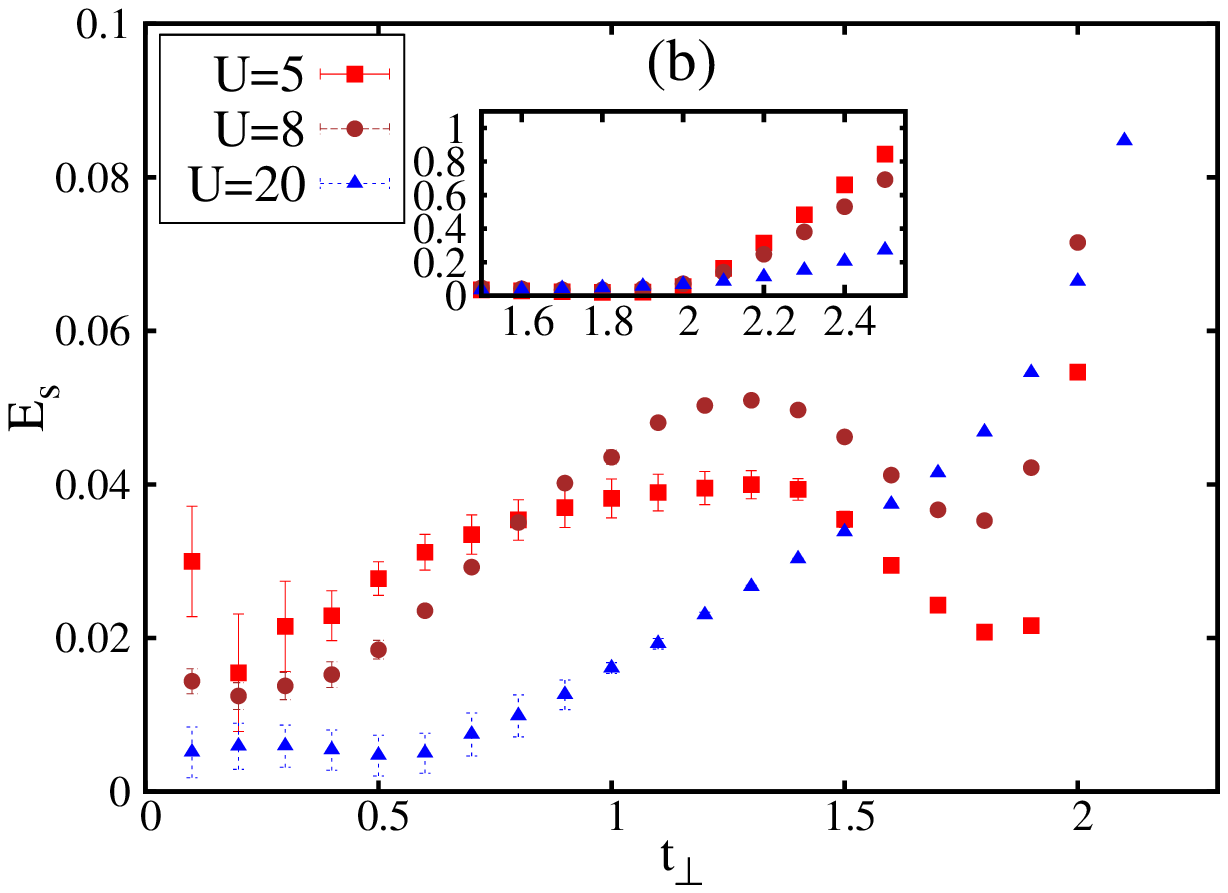}
\caption{\label{fig07} (Color online)
(a) Charge gap $E_{\text{c}}$ and (b) spin gap $E_{\text{s}}$
calculated with the DMRG as a function of the rung hopping $t_{\perp}$ 
in a half-filled asymmetric two-leg Hubbard ladder with $L=128$ rungs.
Finite-size corrections are of the order of the symbol size in (a)
and of the order of $0.01 t_{\parallel}$ in (b).
Error bars indicate DMRG truncation errors larger than the symbol size.
}
\end{figure}

In this section, we discuss the excitation gaps calculated for a half-filled asymmetric ladder.
The gap for charge excitations in a ladder with $N=2M$ electrons is
\begin{eqnarray}
E_{\text{c}} & = & \frac{1}{2} \left [ E_{0}(M+1,M+1)+E_{0}(M-1,M-1) \right . 
\nonumber \\ 
&& \hspace*{1em}\left . -2E_{0}(M,M) \right ]\,,
\end{eqnarray}
where $E_{0}(M_{\uparrow},M_{\downarrow})$ denotes the ground-state
energy of Hamiltonian~(\ref{eq:hamiltonian}) with
$M_{\sigma}$ electrons of spin $\sigma$.
It is the gap seen in the dynamical charge structure
factor, which can be probed by electron-energy-loss
spectroscopy.

Figure~\ref{fig07}(a) shows the behavior of the charge gap 
as a function of the interaction $U$ and 
the rung hopping $t_{\perp}$, which is qualitatively similar to
the Hartree-Fock gap $E_{\text{HF}}$ in Fig.~\ref{fig04}.
A closer investigation reveals four distinct regions: region~(I) for very small $t_{\perp}$, where the gap stays at a finite value 
because of finite-size effects, region~(II) where $E_\text{c}$ increases quadratically with $t_{\perp}$, 
region~(III) at intermediate $t_\perp$ where the gap saturates (or even
decreases), and region~(IV) where $E_\text{g}$ increases rapidly with
$t_{\perp}$ and eventually approaches the value of the band gap~(\ref{eq:band-gap})  as expected (see
Secs.~\ref{sec:weak} and~\ref{sec:dimer}). Region~(II) extends to larger
values of $t_{\perp}$ for a stronger interaction $U$, while the onset of 
region~(IV) shifts from $t_{\perp}=2t_{\parallel}$ to larger values as $U$ increases.

The gap in region~(II) can be well fitted to a function $f(t_{\perp})=a + b
\frac{4t_{\perp}^{2}}{U}$, yielding a slope $b$ that increases from $b\approx 1.1$ for $U=5t_{\parallel}$ to $b\approx 1.5$
for $U=20t_{\parallel}$. The scaling of the charge gap with $t_{\perp}^{2}$
shows that the gap opening  is related to the effective rung exchange coupling 
$J_{\perp}$ discussed in Secs.~\ref{sec:strong} and~\ref{sec:chains}.
The intercept $a$ is negative, suggesting that the charge
gap could close at a small but finite $t_\perp$. The condition
$f(t_{\perp}^{\text{c}})=0$ yields the critical coupling
$t_{\perp}^{\text{c}}(U)$  below
which the charge gap seems to disappear. For instance, we get
$t_{\perp}^{\text{c}}(U=20t_{\parallel}) \approx
0.85 t_{\parallel}$, $t_{\perp}^{\text{c}}(U=8t_{\parallel}) \approx
0.35 t_{\parallel}$, and $t_{\perp}^{\text{c}}(U=5t_{\parallel}) \approx 0.1
t_{\parallel}$.
Region (I) corresponds roughly to the domain $t^{\phantom{c}}_{\perp} <
t_{\perp}^{\text{c}}(U)$.  

To check the finite-size effects we have performed 
calculations for ladder lengths from $L=20$ to $L=200$ and extrapolated
the charge gap to $L \rightarrow \infty$ 
using a quadratic fit in $1/L$. 
$E_{\text{c}}$ remains finite in the thermodynamic
limit for all parameters $U,t_{\perp}>0$, except for region~(I), 
where the charge gap vanishes as $E_{\text{c}}\approx  6 t_{\parallel} / L $.
For comparison, the exact scaling for a half-filled
tight-binding chain is $E_{\text{c}}=  2\pi t_{\parallel} / L$. 
The scaling confirms that added charges (electrons or holes) go
primarily on the Fermi leg and that the interchain hopping $t_{\perp}$
barely affects low-energy charge excitations in the limit of weak $t_{\perp}$,
see Sec.~\ref{sec:chains}.

The spin gap of a ladder with $N=2M$ electrons is
\begin{equation}
E_{\text{s}}=E_{0}(M+1,M-1)-E_{0}(M,M)\,,
\end{equation}
and corresponds to the excitation gap in the dynamical spin structure
factor. It can be measured using inelastic neutron scattering.
Its behavior as a function of $U$ and $t_{\perp}$  is shown
in Fig.~\ref{fig07}(b). We see that it is qualitatively similar
to that of the charge gap, although the difference between 
regions~(II) and~(III) is less clear. 
In addition, for large enough $t_{\perp}$, both gaps approach the value
of the band gap~(\ref{eq:band-gap}), as expected.
For smaller $t_\perp$,  the spin gap is generally (much) smaller
than the charge gap.

Finite-size scaling reveals that the spin gap
is finite in the thermodynamic limit for all parameters $U, t_{\perp}>0$,
except for region~(I), where $E_\text{s}$ vanishes as 
$E_{\text{s}}\approx c\, t_{\parallel} / L $. The values of the prefactor
$c=c_{\text{DMRG}}$ as deduced from our DMRG data agree well with the exact
values $c=c_{\text{BA}}$ obtained from the Bethe ansatz (BA) solution for the 1D Hubbard model on
an open chain~\cite{hubbard-book}. For instance, for moderate interactions
$\left ( U=5t_{\parallel},t_{\perp}=0.1t_{\parallel} \right )$ we get
$c_{\text{DMRG}}\approx c_{\text{BA}} \approx 2.23$, while for $\left
(U=8t_{\parallel},t_{\perp}=0.3t_{\parallel} \right )$
we obtain  $c_{\text{DMRG}}\approx 1.49$ vs. $c_{\text{BA}} \approx 1.51$, and for strong
interactions $ \left ( U=20t_{\parallel},t_{\perp}=0.5t_{\parallel} \right )$ we
find $c_{\text{DMRG}}\approx 0.681$ vs. $c_{\text{BA}} \approx 0.637$. 
This scaling confirms that the lowest triplet excitation 
is essentially a spin excitation of the Hubbard leg and that the interchain hopping 
$t_{\perp}$ barely affects it in the limit of weak $t_{\perp}$,
see Sec.~\ref{sec:chains}.
Moreover, the different prefactors for the finite-size charge and spin gaps are a signature
of the dynamical separation of charge and spin excitations (i.e., different
charge and spin velocities) in the infinite ladder system.

\begin{figure}
\includegraphics[width=0.39\textwidth]{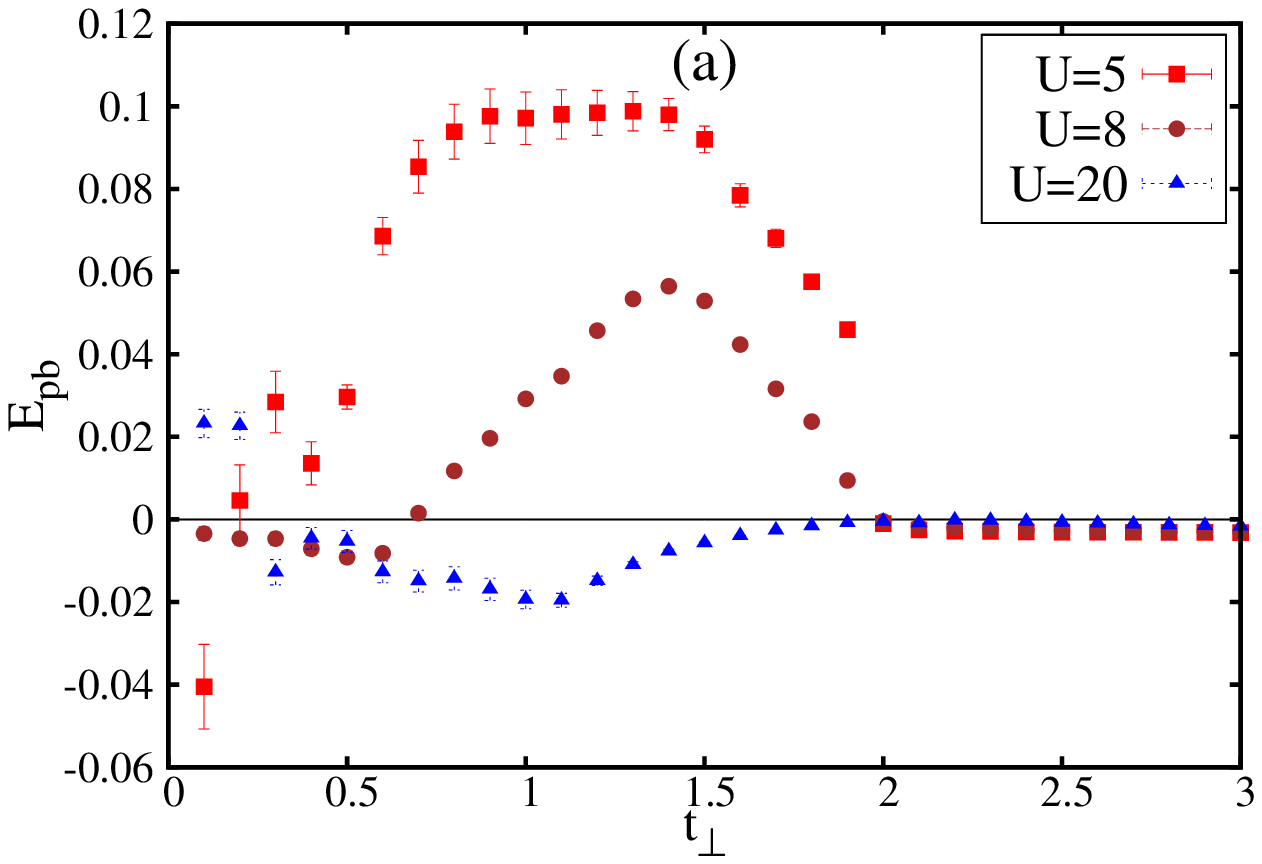}
\includegraphics[width=0.39\textwidth]{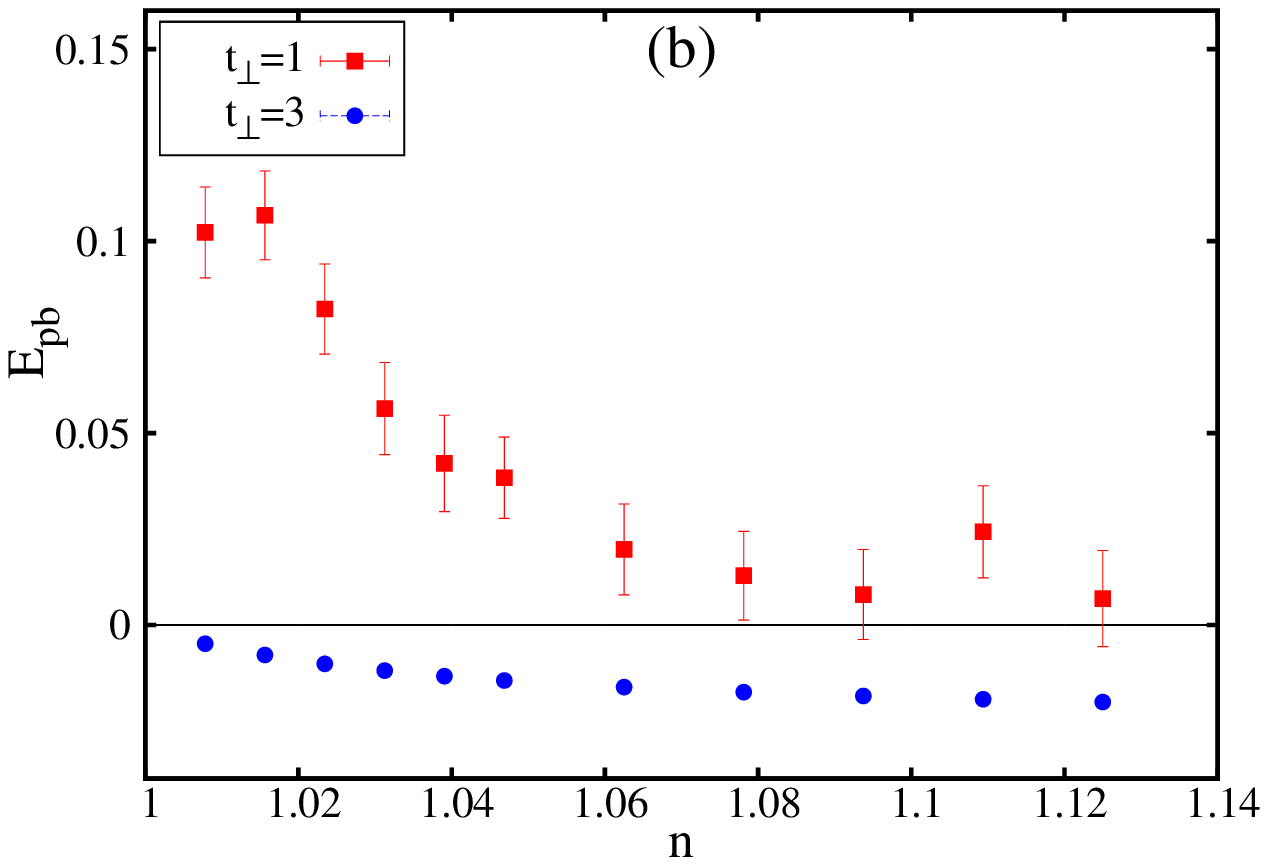}
\caption{\label{fig08} (Color online)
Pair binding energy $E_{\text{pb}}$  calculated with DMRG in an asymmetric two-leg Hubbard ladder with $L=128$ rungs
(a) as a function of the rung hopping $t_{\perp}$ at half-filling for several values of the Hubbard interaction $U$
and (b) as a function of the band filling $n=N/2L$ 
for $U=5t_{\parallel}$ and two values of $t_{\perp}$.
Finite-size corrections are smaller than $0.05t_{\parallel}$.
Error bars indicate DMRG truncation errors larger than the
symbol size. 
}
\end{figure}

The single-particle gap for a ladder with $N=2M$ electrons is defined as
\begin{equation}
\label{eq:sp_gap}
E_{\text{p}}=E_{0}(M+1,M)+ E_{0}(M-1,M)-2E_{0}(M,M)\,.
\end{equation}
This is the gap for the excitations seen in the single-particle spectral
function discussed in Sec.~\ref{sec:qmc} and experimentally accessible
by angle-resolved photoemission spectroscopy. 
We find that $E_\text{p}$ equals the charge gap for weak and strong rung hopping
but differs significantly from it in the intermediate regime.
The difference
\begin{equation}
 E_{\text{pb}}=2(E_{\text{p}}-E_{\text{c}})
\end{equation}
is called the pair binding energy and is shown in Fig.~\ref{fig08}(a).
A significant binding energy only exists for moderate interactions
$5t_{\parallel} \alt U  \alt 8t_{\parallel}$ and intermediate rung hoppings $0.5t_{\parallel} \alt t_{\perp} \alt 2.0t_{\parallel}$.
This corresponds roughly to region~(III) where both charge and spin gaps
saturate or decrease with increasing $t_\perp$.
The study of finite-size effects confirms that $E_\text{pb}$ remains
finite in the limit of infinite ladder length.  
In the other three regions, the pair binding energy is very small or negative
and vanishes in the thermodynamic limit.

It is interesting to study the effect of charges added to the half-filled system.  
Upon doping, the charge and spin gaps close within the accuracy of our calculations 
(limited by finite-size effects and truncation errors). However, the single-particle gap seems to remain
finite at low doping in region~(III) as shown in Fig.~\ref{fig08}(b) for $t_{\perp}=t_{\parallel}$. In the other regions, the pair
binding energy is negligible or even negative, as illustrated in the same figure 
for the case $t_{\perp}=3t_{\parallel}$ that corresponds to
region~(IV). Pairing of added charges also occurs in half-filled symmetric Hubbard
ladders, but with a finite spin gap~\cite{noa94,jec98}.

Our results for the excitation energies, together with the analysis of
limiting cases in Sec.~\ref{sec:model},  seem to suggest 
 the existence of (at least) four distinct phases in
the parameter space $\left ( U>0,t_{\perp}>0 \right )$ of the half-filled asymmetric Hubbard ladder.
In region~(I), i.e., for very small rung hopping $t_{\perp}$, we find gapless charge and spin excitations. 
This corresponds to the Luttinger liquid phase which is expected in the limit of weakly coupled chains, see Sec.~\ref{sec:chains}.
In region~(II), i.e., for moderate $t_{\perp}$ or strong repulsion $U$,
the charge gap increases quadratically with $t_{\perp}$ or, equivalently, linearly with an effective
rung exchange coupling $J_{\perp}$. The spin gap also increases with $t_{\perp}$
but its scaling with $J_{\perp}$ is less clear and it is smaller than the charge gap. 
We identify this phase with the Kondo-Mott insulator defined in Secs.~\ref{sec:strong} and~\ref{sec:chains}. 
In region~(III), i.e., for intermediate values of $t_{\perp}$ and $U$,
both charge and spin gaps are finite but exhibit nonmonotonic behavior with increasing rung hopping.
This phase is characterized by a charge gap much larger than the spin gap,
and by a pair binding energy of the same order of magnitude as the spin gap. This is consistent with a spin-gapped
paramagnetic Mott insulator (similar to the state found in half-filled symmetric Hubbard two-leg ladders~\cite{giamarchi,noa94,bal96,sca97,jec98})
which is expected to exist in the weak-interaction limit of the asymmetric ladder (see Sec.~\ref{sec:weak}).
Finally, in region~(IV), i.e., for large $t_{\perp}$, both charge and spin gaps increase monotonically with
the rung hopping and approach the band gap~(\ref{eq:band-gap}) for large enough $t_{\perp}$.
Region~(IV) corresponds to a correlated band insulator. Indeed, 
the onset of this phase is at $t_{\perp}=2t_{\parallel}$ in the weak-interaction limit (as seen in Sec.~\ref{sec:weak}) and increases
to larger rung hoppings $t_{\perp}$ for stronger interactions $U$, as observed
in the discussion of the dimer limit in Sec.~\ref{sec:dimer}.

Strictly speaking, our DMRG results for the excitation gaps only demonstrate
the existence of two phases (a gapless one and a gapped one) in 
the half-filled asymmetric Hubbard ladder. The distinction between three different insulating phases
has been motivated mainly by the analysis of limiting cases in Sec.~\ref{sec:model} 
and the similarity with the results of the Hartree-Fock approximation in
Sec.~\ref{sec:hartree}. 
In addition, it should be kept in mind that we have not obtained reliable DMRG data when
both the interaction and the rung hopping are small, i.e.,
$U\leq 4t_{\parallel}$ and $t_{\perp}<2t_{\parallel}$. Hence, the distinction
between the three insulating phases remains rather tentative so far.
We now turn to the density profiles of excitations, and later to the single-particle spectral functions,
to demonstrate that the phase diagram indeed includes three qualitatively different gapped phases.

\subsection{Density profiles \label{sec:densities}}

At half-filling, the asymmetric ladder exhibits uniform charge and spin
densities. Other ground-state expectation values such as bond correlations
also show some structure as a result of the open boundary conditions used, but 
we have not found any significant pattern while varying the model parameters $t_{\perp}$ and $U$.
However, we have obtained much information from the charge-- and spin-density variations 
associated with the excitations discussed in the previous sections 
(added electrons/holes and triplet spin excitations).
First of all, the density variations confirm that added charges go primarily on the
Fermi leg while a triplet spin excitation is mostly localized on the
Hubbard leg. This bias becomes larger with stronger interaction $U$ 
but decreases when the rung hopping increases, which is consistent
with our analysis of the various limiting cases in Sec.~\ref{sec:model}.

\begin{figure}
\includegraphics[width=0.39\textwidth]{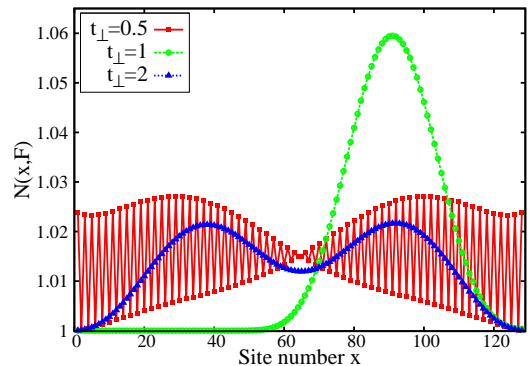}
\caption{\label{fig09} (Color online)
Ground-state charge density distribution on the Fermi leg for two electrons added
to a half-filled ladder with $U=8t_{\parallel}$ and three values of 
$t_{\perp}$. 
}
\end{figure}

The variations of the charge density along the legs also provide us  with useful
information about the different phases. For instance, 
Fig.~\ref{fig09} shows the ground-state charge density on the Fermi leg,
\begin{equation}
N(x,F) = \left \langle \psi \left \vert n_{x,F} \right \vert \psi \right \rangle,
\end{equation}
where $n_{x,y} = n_{x,y,\uparrow} + n_{x,y,\downarrow}$ and $\vert \psi \rangle$ denotes the ground state,
when two electrons are added to a half-filled ladder with $U=8t_{\parallel}$.
We clearly see three qualitatively different density profiles. 
In the Kondo-Mott insulator phase $\left ( t_{\perp}=0.5t_{\parallel} \right)$, the
density distribution of the added charges oscillates strongly from one site to the next. 
(Similar patterns exist in the Luttinger liquid phase but the results are less clear-cut
because of larger DMRG errors.)

 In the spin-gapped Mott insulator
phase $\left ( t_{\perp}=t_{\parallel} \right)$, both added charges are
concentrated in a single wave packet on one side of the system as if they were bound together. 
This confirms the tendency to binding added charges revealed by the pair binding energy in the previous section.
This charge distribution breaks the reflection
symmetry around the center of the Fermi leg, which indicates that odd and even excitations are degenerate,
at least within the accuracy of our DMRG calculation.
We also observe spin and charge densities that break the reflection symmetry if a single electron is added to the half-filled
ladder. In that case, the symmetry breaking is readily explained by the degeneracy of the lowest
single-particle excitations with wave numbers $k_g$ and $k'_g$, see Sec.~\ref{sec:qmc}.
In an open chain with an even number of sites, the condition $k_g+k'_g=\pi$ (see Sec.~\ref{sec:weak})
implies that one of this state is even while the other one is odd with respect to a reflection.
Thus the DMRG algorithm may return any (symmetry-breaking) linear combination of these two states
for the ground state. We think that a similar (quasi-) degeneracy occurs for two-particle excitations.
(We have also investigated the ground state with up to 32 electrons added to a half-filled $2\times128$
ladder and found no sign of phase separation.)


In the correlated band insulator phase $\left ( t_{\perp}=2t_{\parallel} \right)$, 
the added charges appear to be independent. Actually, their density distribution 
corresponds to two free particles in a tight-binding box.
In conclusion, the distinct density profiles for added charges confirm the existence of three different gapped
phases and the tendency for pair binding in the spin-gapped Mott insulating phase.

\begin{figure}
\includegraphics[width=0.39\textwidth]{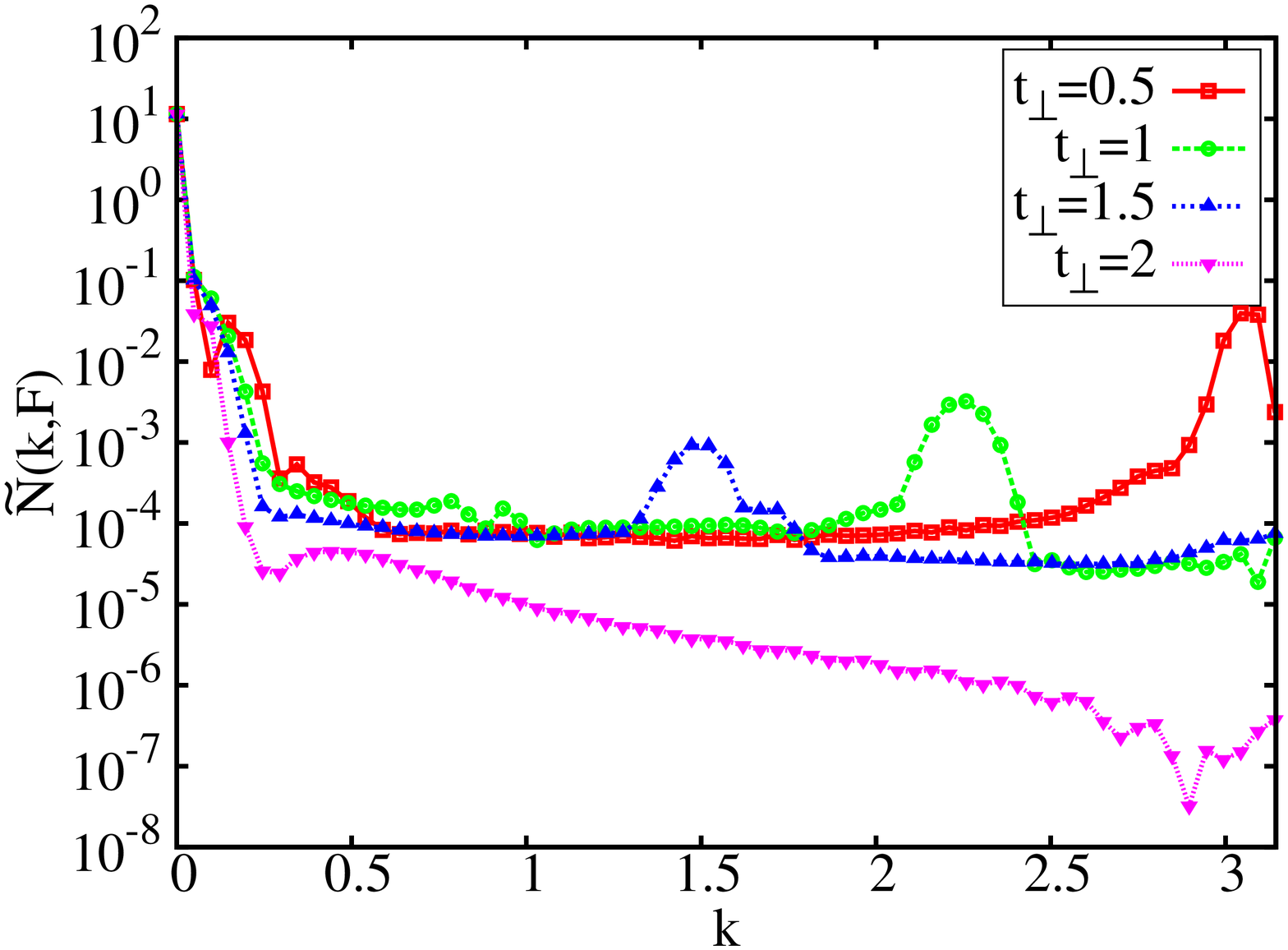}
\caption{\label{fig10} (Color online)
Fourier transform of the ground-state charge density on the Fermi leg 
with two electrons added to a half-filled ladder with $U=5t_{\parallel}$
and four values of $t_{\perp}$. 
}
\end{figure}

A more quantitative study can be made using the Fourier transform of these
density distributions.
For instance, Fig.~\ref{fig10} shows the Fourier transform of the charge
density on the Fermi leg, 
\begin{equation}
\tilde{N}(k,F) = \frac{1}{\sqrt{L}} \left \vert \sum_{x=1}^{L} N(x,F) \exp(-ikx) \right \vert 
\end{equation}
for $k=2\pi z/L$ with integers $\vert z \vert < L/2$,
when two electrons are added to a half-filled ladder with $U=5t_{\parallel}$.
The strong peak around $k=0$ is mostly due to the uniform 
density of the half-filled system. If the lowest elementary single-charge excitations have wave numbers
$\pm k_{\text{g}}$ then $\tilde{N}(k,F)$ should exhibit peaks at $k=\pm 2k_{\text{g}} \mod 2\pi$.
We see in Fig.~\ref{fig10} that the residual spectral weight
is concentrated close to $k=\pi$  for the Kondo-Mott insulator ($t_{\perp}=0.5t_{\parallel}$). 
This implies that the lowest excitations have a wave number
$k_{\text{g}} = \pi/2$ in this phase.
In the spin-gapped Mott insulator 
(cf. data for $t_{\perp}=t_{\parallel}$ and $1.5 t_{\parallel}$ in Fig.~\ref{fig10}) 
the spectral weight exhibits peaks at wave numbers $0< \vert q \vert <\pi$. 
This suggests that the low-energy excitations have incommensurate
wave numbers $k_{\text{g}} = \vert q \vert /2$ and $k'_{\text{g}} = \pi - \vert q \vert
/2$ in that phase. 
Finally, in the correlated band insulator phase ($t_{\perp}=2t_{\parallel}$ in Fig.~\ref{fig10}) we observe no other
structure than the $k=0$ peak. This corresponds to low-energy excitations
with wave numbers $k_{\text{g}} = 0$ or $\pi$.

Similarly, we have studied the spin distribution of the lowest triplet eigenstate as well
as the charge and spin distributions for one added electron.
All results are compatible with the above analysis:
low-energy single-particle excitations have wave numbers $\pm \pi/2$ in 
the Luttinger liquid and Kondo-Mott insulator, incommensurate wave numbers in the spin-gapped Mott insulator,
and wave numbers $0$ or $\pi$ in the correlated band insulator.
These results also agree perfectly with the analysis of the limiting cases in
Sec.~\ref{sec:model}.

Somewhat surprisingly, the presence of three gapped phases with distinct 
low-energy excitations is correctly predicted by the Hartree-Fock
approximation, see Sec.~\ref{sec:hartree}.
However, the latter is otherwise quite inaccurate as it 
predicts an antiferromagnetic Mott insulator or a band insulator with
antiferromagnetic long-range order for all parameters
$ U, t_{\perp} > 0$, while the (almost exact) DMRG results confirm the absence of any
antiferromagnetic long-range order (and also reveal the existence of an additional, gapless phase).

\subsection{Correlation functions}

The DMRG method has been used to compute static correlation functions of ladder 
systems~\cite{dmrg2,dmrg3,noa94,rob12}.
Unfortunately, their interpretation can be rather difficult because of the open boundary
conditions. In the asymmetric Hubbard ladder~(\ref{eq:hamiltonian}), it is
further complicated by the different behavior of the two legs. Nevertheless,
we calculated, e.g., charge--charge and spin--spin correlations as well as various singlet and triplet
pairing correlations. 
Typically, we can obtain accurate results for small system lengths $L$, or for
short distances $x$, but long-distance correlations
are quite inaccurate because of an insufficient DMRG convergence. Thus we have not succeeded
in gaining much useful information for the asymptotic behavior of correlation functions. 

In the Luttinger liquid phase, we find dominant antiferromagnetic spin correlations with
a power-law decay $x^{\alpha}$ and exponents $\alpha$ close to $-1$, as in
a half-filled Hubbard chain.
In the correlated band insulator phase, with its large charge and spin gaps, we
observe that all correlations decay exponentially.  
In the two other phases (Kondo-Mott and spin-gapped Mott insulators), however, 
we find a rapid (faster than $x^{-2}$) but apparently nonexponential decay of correlation functions.
Clearly, in those cases, the correlation lengths are larger than our system sizes (up to
$L=128$ rungs) and we do not see the asymptotic behavior.

We also investigated correlation functions of the asymmetric Hubbard ladder away from half-filling 
to understand the nature of the charge pairing observed when electrons or holes 
are added to a half-filled ladder in the spin-gapped Mott insulating phase.
Unfortunately, we do not find any enhanced pairing correlations and hence do not understand the
structure of these pairs.
Among all the pairing correlation functions that we examined, pair-density-wave (PDW) correlations~\cite{ber10}
decrease most slowly.
 PDW correlations in two-leg ladder systems have attracted much interest recently~\cite{ber10,alm10,jae12,rob12,dob13}
because they resemble correlations in the PDW state which was proposed to describe the phenomenology of stripe-ordered
high-temperature superconductors. Interestingly,
dominant quasi-long range PDW correlations have been found in a spin-gapped phase
of the Kondo-Heisenberg model away from half-filling~\cite{ber10}.
In the asymmetric Hubbard ladder close to half-filling, however, we find that  PDW correlation
functions decay as $x^{-2}$ or faster with distance $x$.
The dominant correlations seem to be power-law charge and spin correlations with
exponents $\alpha$ between $-1$ and $-2$.
For comparison, in the symmetric Hubbard ladder close to half-filling,
the dominant pairing correlations are of the $d$-wave type but they are not
enhanced, i.e., they decay as $x^{-2}$ like for a noninteracting ladder
($U=0$) \cite{noa94,jec98}.

\section{\label{sec:qmc} Spectral functions}

Our analysis of excitation density profiles in Sec.~\ref{sec:densities} and
the Hartree-Fock approximation
in Sec.~\ref{sec:hartree} 
suggests that the lowest elementary excitations have different wave numbers 
$k_{\text{g}}$ in the three gapped phases that exist at half-filling.
To confirm this hypothesis, we consider the momentum and energy-resolved
single-particle spectral function, which can be probed experimentally 
using angle-resolved photoemission spectroscopy. The sharp maxima at
the spectrum onset in correlated  electron systems~\cite{tsv11,ben04,jec08}
allows us to determine $k_{\text{g}}$.

Although the single-particle spectral function can in principle be calculated with the
DMRG method~\cite{ben04,jec08}, such calculations come at a high computational cost
and  the interpretation of the results is complicated by the use of
pseudo-wave numbers for open boundary conditions.
(For instance, we can see in Fig.~\ref{fig10} that peaks of a Fourier spectrum are still
considerably smeared by boundary effects even for large ladders with 128 rungs.)
 Instead, we calculate the
spectral function using the CT-INT continuous-time quantum Monte Carlo
method~\cite{rub05}, which is based on a weak-coupling expansion in the interaction
$U$, and gives exact results for finite systems and finite temperatures.
A detailed review of the method has been given in Ref.~\cite{gull11}. We used
single-vertex updates and Ising spin flips, and simulated ladders with 
periodic boundary conditions along the legs. 

With the help of the stochastic maximum entropy method \cite{bea04}, we
can perform the necessary analytic continuation of the QMC results for the single-particle Green function
$G(k,y,\tau)=\langle c^{\dag}_{k,y,\sigma} (\tau) c^{\phantom{\dag}}_{k,y,\sigma}(0) \rangle$
to obtain the single-particle spectral function
\begin{align}\label{eq:akw}\nonumber
  A(k,y,\omega)
  &=
  \frac{1}{Z}\sum_{ij}
  {|\langle {i}| c_{k,y,\sigma} |{j}\rangle|}^2 (e^{-\beta E_i}+e^{-\beta E_j})
  \\
  &\hspace*{4em}\times
  \delta(\Delta_{ji}-\omega)
  \,.
\end{align}
Here, $c_{k,y,\sigma}$ is the Fourier transform of
$c_{x,y,\sigma}$ in the leg direction, $Z$
is the grand-canonical partition function, 
$|{i}\rangle$ is an eigenstate with energy $E_i$, and $\Delta_{ji}=E_j-E_i$.
We carried out simulations for closed-shell configurations ($L=30$)
and open-shell configurations ($L=32$) at inverse temperatures $\beta
t_{\parallel}=30$ and $32$, respectively. 
We did not observe any significant finite-size effect for the wave number of
the lowest excitations. The analytical continuation introduces some quantitative
uncertainties, but the overall features of the spectral functions
are robust and fully agree with the results obtained above. Because
closed-shell results are usually more reliable and more representative of the
thermodynamic limit, we only report the latter below.

QMC methods were used to study spectral functions of symmetric ladders
in Refs.~\cite{dah97,sca97,sca02}. Because symmetric ladders conserve
the parity under reflection in the rung direction, the spectral function was investigated separately
for the bonding and antibonding orbitals.  For the asymmetric ladder studied
here, it is more convenient to consider the spectral function for the Hubbard
and Fermi legs separately, as indicated by $y$ in Eq.~(\ref{eq:akw}).
As a result of the particle-hole symmetry of
Hamiltonian~(\ref{eq:hamiltonian})  at half-filling, $A(k,y,\omega)$ has the
symmetry property $A(k,y,-\omega) = A(k+\pi,y,\omega)$. Consequently, the
single-particle gap is symmetric around $\omega=0$. In addition, the system is symmetric under a reflection in the
leg direction and thus $A(-k,y,\omega)=A(k,y,\omega)$.

\begin{figure}
\includegraphics[width=0.49\textwidth]{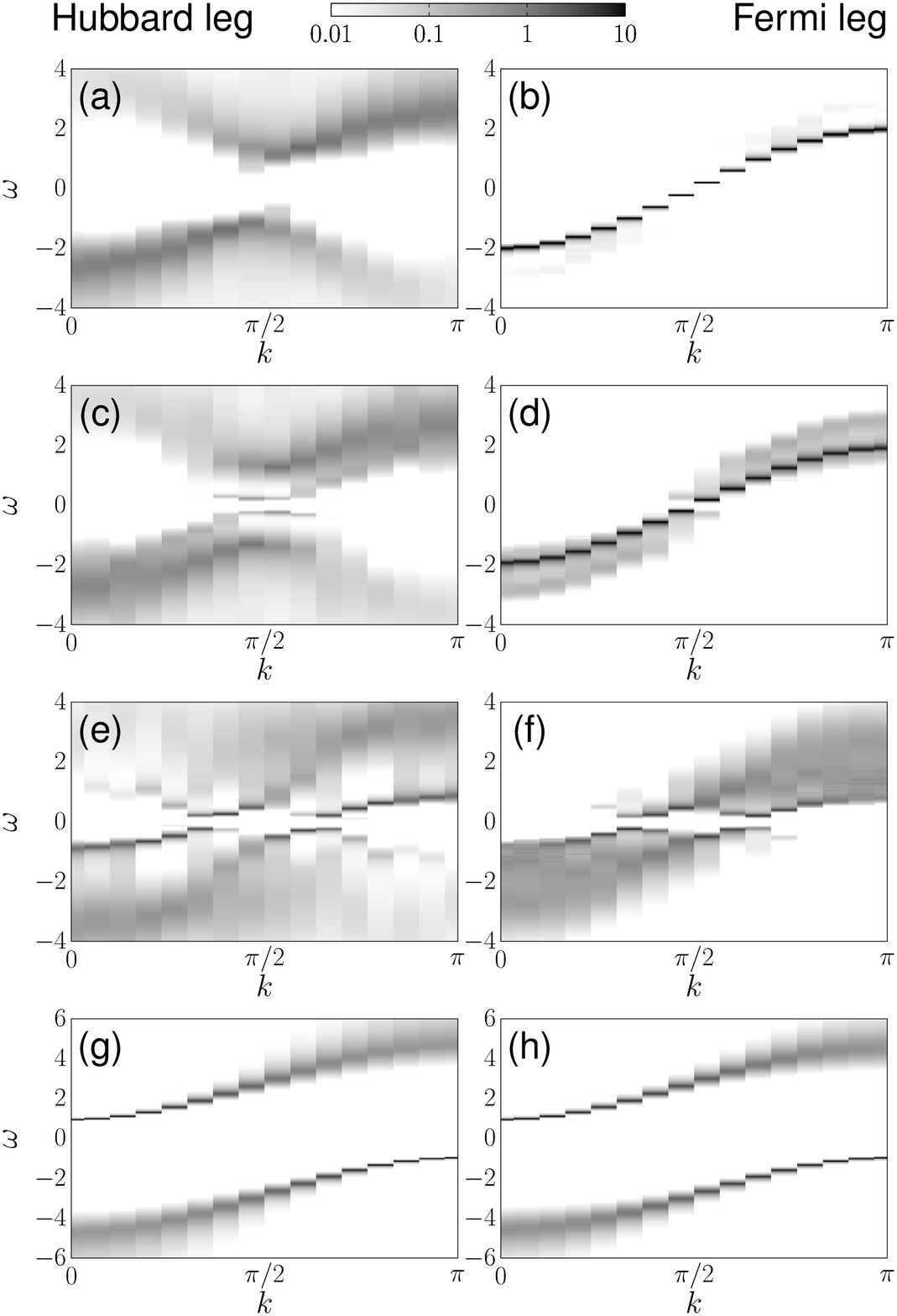}
\caption{\label{fig11} 
Spectral functions $A(k,y,\omega)$ on the Hubbard leg (left column) and the
Fermi leg (right column) calculated using the CT-INT method with $\beta t_{\parallel}=30$ on a
periodic ladder with $L=30$ rungs and  $U=5t_{\perp}$. (a),(b) Luttinger
liquid phase ($t_{\perp}=0.1 t_{\parallel}$), (c),(d) Kondo-Mott insulator
($t_{\perp}=0.3 t_{\parallel}$), (e),(f) incommensurate spin-gapped Mott insulator ($t_{\perp}=t_{\parallel}$),
(g),(h) correlated band insulator ($t_{\perp}=3t_{\parallel}$).
}
\end{figure}

The spectral functions for the Hubbard and Fermi legs in the four different
phases of the model~(\ref{eq:hamiltonian}) are shown in Fig.~\ref{fig11}. The
interaction is fixed to $U=5t_\parallel$, while the hopping $t_\perp$
increases from top to bottom, leading to a progression from 
weakly coupled chains to a true ladder system with strong rung hopping.

In the Luttinger liquid phase, Figs.~\ref{fig11}(a) and (b), the spectrum on
the Hubbard leg looks clearly different from the free-particle like spectrum
on the Fermi leg. There is substantial weight at $\omega=0$ for the
Fermi wave number $k_\text{F}=\pi/2$, indicating metallic behavior.
Away from $\omega=0$ the main spectral features still reflect
the dispersion of elementary excitations in independent chains, 
compare with Fig.~\ref{fig02}(c).

For the Kondo-Mott insulator phase [see Figs.~\ref{fig11}(c) and (d)]
the lowest excitations are clearly located at $k_{\text{g}}=\pi/2$.
The gap is not visible because the true gap expected from the DMRG
calculations is only a pseudogap as a result of the finite temperature
used in the CT-INT simulations. Nevertheless, all results in
Fig.~\ref{fig11} are compatible with our findings for the DMRG
single-particle gap~(\ref{eq:sp_gap}). The spectral function of the Hubbard
leg in Fig.~\ref{fig11}(c) resembles that of a Hubbard
chain~\cite{ben04,jec08}  while the spectral function of the Fermi leg
[Fig.~\ref{fig11}(d)] looks
quite similar to Fig.~\ref{fig11}(b) but with signs of the pseudogap at
$\omega=0$, $k=\pi/2$.

For the spin-gapped Mott phase we see in Figs.~\ref{fig11}(e) and (f)
that the lowest excitations are at wave numbers $k_{\text{g}}$
and $k'_{\text{g}}$, which are quite symmetrically located around $\pi/2$,
so that $k_{\text{g}} + k'_{\text{g}} \approx \pi$.
Thus, in this intermediate regime of $t_{\perp}$,
the lowest single-particle excitations have incommensurate wave numbers.  
Incommensurability in the excitation spectrum has also been found in the
half-filled symmetric Hubbard ladder with moderate rung hopping~\cite{noa94},
in a frustrated Kondo-Heisenberg model~\cite{eid11}, and in various
correlated 1D systems such as the bilinear biquadratic spin-1 chain~\cite{gol99}
and a two-leg spin ladder with nearest and next-nearest coupling~\cite{lav11,shy13}.
In contrast to the DMRG, the CT-INT method also yields accurate results for
weak on-site repulsion $U$, and shows that an incommensurate
excitation spectrum  exists down to at least $U=3t_{\parallel}$ for $t_{\perp}=t_{\parallel}$. 
We suspect that this phase remains as $U\rightarrow0$ and could be investigated
with field-theoretical approaches starting from a noninteracting 
asymmetric ladder, as discussed in Sec~\ref{sec:weak}. 

Finally, in the correlated band insulator regime shown in
Figs.~\ref{fig11}(g) and (h), the lowest excitations have
wave number $k_{\text{g}}=\pi$ for particle removal and 
$k_{\text{g}}=0$ for particle addition, respectively.
The spectra are almost identical on the two legs. This agrees with the analysis
of the weak-interaction limit in Sec.~\ref{sec:weak} [compare with
Fig.~\ref{fig02}(a)] and  the dimer limit in Sec.~\ref{sec:dimer}. Indeed,
when $t_{\perp}$ is large enough, elementary excitations become almost
(anti-)symmetric with respect to a
reflection in the rung direction. Obviously, this case is very similar to 
a half-filled symmetric Hubbard ladder with a strong rung hopping. 

The markedly distinct spectral functions in Fig.~\ref{fig11} confirm the
existence of one metallic and three 
different gapped phases in the asymmetric Hubbard ladder at half-filling.
The phases can be characterized by the wave numbers of the low-energy excitations,
in agreement with the analysis of limiting cases, the Hartree-Fock approximation, and the
DMRG density profiles.

\section{Conclusions}\label{sec:conc}

\begin{figure}
\includegraphics[width=0.39\textwidth]{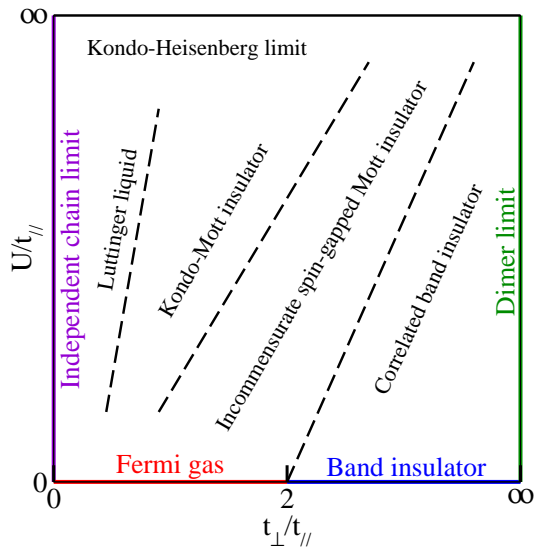}
\caption{\label{fig12} 
(Color online) Schematic phase diagram of the half-filled asymmetric Hubbard ladder.
}
\end{figure}

In this work, we studied the rich physics of the half-filled asymmetric
ladder model~(\ref{eq:hamiltonian}). In particular,  we found 
three gapped phases that differ in the shape of their
single-particle excitation spectra, in addition to a Luttinger liquid phase.
For  strong Hubbard interaction $U$ or weak interchain hopping $t_{\perp}$, 
our model is related to the Kondo-Heisenberg model, whereas
for weak Hubbard repulsion $U$ or strong rung hopping $t_{\perp}$, 
it is similar to that of a half-filled symmetric Hubbard ladder.
Although we do not have enough data to draw a quantitative phase diagram,
we show in Fig.~\ref{fig12} a schematic and tentative phase diagram that
summarizes our findings.
Surprisingly, the overall structure is similar to the Hartree-Fock ``phase diagram'' in Fig.~\ref{fig06}
including, in particular, the wave numbers of the lowest single-particle excitations. 
The main differences are the presence of a Luttinger liquid phase 
at small interchain hopping and the absence of long-range antiferromagnetic order.

The three gapped phases are not  differentiated by  a symmetry breaking or
a gap closing but only by a change of the wave number of the low-energy excitations.
Similar transitions between phases with commensurate and incommensurate
low-energy excitations were found previously in other models, such as the bilinear-biquadratic spin-1 chain~\cite{gol99}.
It is difficult to determine phase boundaries numerically for phase
transitions that do not involve any symmetry breaking or gap closing.
In recent years, various measures of entanglement have been proposed
as useful tools for the study of quantum phase transitions~\cite{ost02,wu04,leg06,leg07,mun09}.
We examined one of them, the block entropy in the middle
of the lattice, using the DMRG method. Although we observed a different scaling of this entropy with block size in the gapless 
phase compared to the gapped ones, we did not found any feature which
could help locate the boundaries between the three gapped phases.
Nevertheless, it is likely that DMRG calculations combined with one of the more sophisticated entanglement-based methods 
could provide a more precise phase diagram.

The existence of a Luttinger liquid phase has been demonstrated within the accuracy of 
our numerical methods.
It should be kept in mind, however, that exponentially small energy scales usually 
associated with Kondo physics are not accessible with these methods. Therefore, we cannot rigorously 
exclude the existence of other phases with exponentially small gaps in the limit of very 
small interchain hopping. 
We think that  the best approach to solve this issue,  and more generally to improve our understanding of the
asymmetric Hubbard ladder, is a more systematic investigation of the limiting cases in Sec.~\ref{sec:model}.
On the one hand, effective models for the low-energy physics can be derived in the strong-interaction $(U \gg t_{\parallel})$
and dimer $(t_{\perp} \gg t_{\parallel})$ limits. They should be more amenable to our numerical methods and 
simple analytical approximations and could thus provide us with a better understanding of the upper and right-hand-side parts of
the phase diagram in Fig.~\ref{fig12}.
On the other hand,  it is likely that field-theoretical methods for weakly-coupled chains (see Sec.~\ref{sec:chains}) and 
weakly-interacting ladders (Sec.~\ref{sec:weak}) could be used to investigate 
the left-hand and lower parts of the phase diagram. 

This study was motivated by the problem of correlated quantum wires deposited on a substrate.
In this context, our results confirm that 1D correlated systems are extremely sensitive to their environment.
Their properties can be drastically modified by   
varying the strength of the hybridization (the hopping $t_{\perp}$) between the interacting wire (the Hubbard leg) and
the noninteracting substrate (the Fermi leg).
In that perspective, the study of asymmetric ladder models constitutes a useful approach for exploring
the basic physics of a quantum wire deposited on a substrate.

Yet we also face some problems with this approach.
Clearly, it is not enough to represent the substrate by a single chain because
the wire interaction can then dominate the full system as our results show. 
Instead, the substrate should include many more explicit degrees of freedom than the wire.
This could be realized using wider ladders with several legs representing the substrate.
Indeed, it is possible to map the Hamiltonian of some wire-substrate systems exactly onto  ladder models
with an infinite number of inequivalent legs.  (A similar idea has been recently
used to map multiple multi-orbital impurities on a honeycomb lattice onto 
effective multi-leg ladder systems~\cite{shi14}.) An effective ladder model with $n+1$ legs
can then be seen as the ``$n$-th order'' approximation of the substrate degrees of freedom.
We think that this approach could enable a more systematic study of wire-substrate systems
in the future.

In addition, in most experiments, the substrate is a band insulator. This condition can be easily realized using
two or more orbitals per site but this will double the number of model parameters (at least).
This reveals the most serious practical difficulty: we do not know which parameter
regime  is appropriate for real systems such as atomic wires deposited on substrates.
Therefore,  ladder models cannot currently be used to study specific materials but can only
provide generic information about the physics of quasi-1D electron systems.
However, we think that systematic studies of effective $n$-leg ladder models
could enable the determination of appropriate model parameters by comparison with experiments
and first-principles simulations for wire-substrate systems.

\begin{acknowledgments}
We thank R. M. Noack and A. Rosch for helpful discussions. This work has been done
as part of the Research Units \textit{Metallic nanowires on the atomic scale: Electronic
and vibrational coupling in real world systems} (FOR1700) and 
\textit{Advanced Computational Methods for Strongly Correlated Quantum Systems}
(FOR1807) of the German Research Foundation (DFG) and was supported by
grants Nos.~JE~261/1-1
and Ho~4489/2-1. The DMRG calculations were carried out on the cluster system
at the Leibniz University of Hannover and at the Sudan Center for HPC and
Grid Computing. The QMC simulations were performed at the J\"ulich Supercomputing Centre.
\end{acknowledgments}


\begin{thebibliography}{99}
\bibitem{giamarchi}  T. Giamarchi, \textit{Quantum Physics in One Dimension} (Oxford University Press, Oxford, 2007).
\bibitem{noa94} R.~M. Noack, S.~R. White, and D.~J. Scalapino, 
Phys. Rev. Lett. \textbf{73}, 882 (1994).
\bibitem{bal96} L. Balents and M.~P.~A. Fisher, Phys. Rev. B \textbf{53}, 12133 (1996).
\bibitem{sca97} D.~J. Scalapino, Physica C \textbf{282-287}, 157 (1997). 
\bibitem{jec98} E. Jeckelmann, D.~J. Scalapino, and S.~R. White, Phys. Rev. B \textbf{58}, 9492 (1998).
\bibitem{con05} D. Controzzi and A.~M. Tsvelik, Phys. Rev. B \textbf{72}, 035110 (2005).
\bibitem{tsv11} A.~M. Tsvelik, Phys. Rev. B \textbf{83}, 104405 (2011).
\bibitem{rob12} N.~J. Robinson, F.~H.~L. Essler, E. Jeckelmann, and A.~M.  
Tsvelik, Phys. Rev. B \textbf{85}, 195103 (2012).
\bibitem{car13} S.~T. Carr, B.~N. Narozhny, and A.~A. Nersesyan,
Annals of Physics \textbf{339}, 22 (2013).
\bibitem{whi94} S.~R. White, R.~M. Noack, and D.~J. Scalapino, 
Phys. Rev. Lett. \textbf{73}, 886 (1994).
\bibitem{dah97} T. Dahm and D.~J. Scalapino, Physica C \textbf{288}, 33 (1997).
\bibitem{sca02} D.~J. Scalapino, Physica B \textbf{318}, 92 (2002).
\bibitem{lav11} A. Lavar\'{e}lo, G. Roux, and N. Laflorencie, Phys. Rev. B \textbf{84} 144407 (2011).
\bibitem{shy13} I.~T. Shyiko, I.~P. McCulloch, J.~V. Gumenjuk-Sichevska, and A.~K. Kolezhuk, Phys. Rev. B~\textbf{88}, 014403 (2013).
\bibitem{sik97} A. E. Sikkema, I. Affleck, and S. R. White, Phys. Rev. Lett.  \textbf{79}, 929 (1997).
\bibitem{zac01b} O. Zachar and A. M. Tsvelik, Phys. Rev. B \textbf{64}, 033103 (2001).
\bibitem{ber10} E. Berg, E. Fradkin, and S. A. Kivelson, Phys. Rev. Lett.  \textbf{105}, 146403 (2010).
\bibitem{dob13} A. Dobry, A. Jaefari, and E. Fradkin, Phys. Rev. B \textbf{87}, 245102 (2013).
\bibitem{eid11} E. Eidelstein, S. Moukouri, and A. Schiller, Phys. Rev. B \textbf{84} 014413 (2011).
\bibitem{alh09} K. A. Al-Hassanieh, C. D. Batista, P. Sengupta, and A. E.  Feiguin, Phys. Rev. B \textbf{80}, 115116 (2009).
\bibitem{spr07} M. Springborg and Y. Dong, \textit{Metallic Chains / Chains of Metals} (Elsevier, Amsterdam, 2007). 
\bibitem{das01} I.~K. Dash and A.~J. Fisher, J. Phys.: Condens. Matter \textbf{13}, 5035 (2001).
\bibitem{bae04} D. Baeriswyl and L. Degiorgi (Eds.), \textit{Strong Interactions in Low Dimensions}
(Kluwer Academic Publishers, Dordrecht, 2004).
\bibitem{kag82} S. Kagoshima, H. Nagasawa, and T. Sambongi,
\textit{One-Dimensional Conductors} (Springer, Berlin, 1982).
\bibitem{gruener} G. Gr\"{u}ner, \textit{Density Waves in Solids} (Perseus
Publishing, Cambridge, 2000).
\bibitem{bechgaard} M. Dressel, ISRN Condens. Matter Phys. \textbf{2012}, 732973 (2012).
\bibitem{kiess} H. Kiess (ed.), \textit{Conjugated Conducting Polymers}
(Springer, Berlin, 1992).
\bibitem{onc08} N. Oncel, J. Phys.: Condens. Matter \textbf{20}, 393001 (2008).
\bibitem{sni10} P.~C. Snijders and H.~H. Weitering, Rev. Mod. Phys. \textbf{82}, 307 (2010).
\bibitem{blu11} C. Blumenstein, J. Sch\"{a}fer, S. Mietke, S. Meyer, 
A. Dollinger, M. Lochner, X.~Y. Cui, L. Patthey, R. Matzdorf, and R. Claessen, 
Nature Physics \textbf{7}, 776 (2011). 
\bibitem{dmrg} S. R. White, Phys. Rev. Lett. \textbf{69}, 2863 (1992); Phys.  Rev. B \textbf{48}, 10345 (1993).
\bibitem{dmrg2} U. Schollw\"{o}ck, Rev. Mod. Phys. \textbf{77}, 259 (2005).
\bibitem{dmrg3} E. Jeckelmann, in \textit{Computational Many Particle Physics} (Lecture Notes in Physics \textbf{739}),
edited by H. Fehske, R. Schneider, and A. Wei\ss e (Springer-Verlag, Berlin, Heidelberg, 2008), p. 597.
\bibitem{rub05} A. N. Rubtsov, V. V. Savkin, and A. I. Lichtenstein,
  Phys. Rev. B {\bf 72}, 035122 (2005).
\bibitem{sch66} J.~R.~Schrieffer and P.~A.~Wolff, Phys. Rev. \textbf{149}, 491 (1966).
\bibitem{lac79} C. Lacroix and M. Cyrot, Phys. Rev. B \textbf{20}, 1969 (1979).
\bibitem{zac01a} O. Zachar, Phys. Rev. B \textbf{63}, 205104 (2001).
\bibitem{esk94} H. Eskes, A.~M. Ole\'s, M.~B.~J.~Meinders, and W. Stephan, Phys.
Rev. B \textbf{50}, 17980 (1994).
\bibitem{hubbard-book} F.H.L. Essler, H. Frahm, F. G\"ohmann, A. Kl\"umper, and
V. Korepin, \textit{The One-Dimensional Hubbard Model} (Cambridge University
Press, Cambridge, 2005).
\bibitem{gebhard} F.~Gebhard, \textit{The Mott Metal-Insulator Transition}
(Sprin\-ger, Berlin, 1997).
\bibitem{alm10} J. Almeida, G. Roux, and D. Poilblanc, Phys. Rev. B~\textbf{82}, 041102 (2010).
\bibitem{jae12} A. Jaefari and E. Fradkin, Phys. Rev. B~\textbf{85}, 035104 (2012).
\bibitem{ben04} H. Benthien, F. Gebhard, and E. Jeckelmann, Phys. Rev. Lett. \textbf{92}, 256401 (2004). 
\bibitem{jec08} E. Jeckelmann, Progress of Theoretical Physics Supplement \textbf{176}, 143 (2008). 
\bibitem{gull11} E. Gull, A. J. Millis, A. I. Lichtenstein, A. N. Rubtsov,
  M. Troyer, and P. Werner, Rev. Mod. Phys. {\bf 83}, 349 (2011).
\bibitem{bea04} K. S. D. Beach, arXiv:cond-mat/0403055 (2004).
\bibitem{gol99} O. Golinelli, Th. Jolic\oe ur, and E. S. S\o rensen, Eur. Phys.  J. B \textbf{11}, 199 (1999).
\bibitem{ost02} A. Osterloh, L. Amico, G. Falci, and R. Fazio, Nature \textbf{416}, 608 (2002). 
\bibitem{wu04} L.-A. Wu, M. S. Sarandy, and D. A. Lidar, Phys. Rev. Lett.  \textbf{93}, 250404 (2004).
\bibitem{leg06} \"{O}. Legeza  and J. S\'{o}lyom, Phys. Rev. Lett. \textbf{96}, 116401 (2006).
\bibitem{leg07} \"{O}. Legeza, J. S\'{o}lyom, L. Tincani, and R. M. Noack,  Phys. Rev.  Lett. \textbf{99}, 087203 (2007).
\bibitem{mun09} C. Mund, \"{O}. Legeza, and  R. M. Noack,  Phys. Rev. B \textbf{79}, 245130 (2009).
\bibitem{shi14} T. Shirakawa and S. Yunoki, Phys. Rev. B \textbf{90}, 195109 (2014).
\end{thebibliography}
\end{document}